\documentclass[aps,prb,preprint,floatfix,superscriptaddress,titlepage]{revtex4-2}

\usepackage{graphicx}
\usepackage{amsmath,amssymb,bm}
\usepackage{xcolor}
\usepackage{multirow}
\usepackage{booktabs}
\usepackage{hyperref}

\makeatletter
\providecommand{\@LN}[2]{}
\makeatother

\begin{document}

\title{Computational symmetry hierarchy of time-reversal-even and -odd spin Hall conductivity tensors in altermagnets}

\author{Dameul Jeong}
\affiliation{Department of Physics and Research Institute for Basic Sciences, Kyung Hee University, Seoul 02447, Korea}

\author{Seoung-Hun Kang}
\email{shkang@kisti.re.kr}
\affiliation{Department of Physics and Research Institute for Basic Sciences, Kyung Hee University, Seoul 02447, Korea}
\affiliation{Department of Information Display, Kyung Hee University, Seoul 02447, Korea}
\affiliation{Research Center for Technology Commercialization, Korea Institute of Science and Technology Information (KISTI), Seoul 02456, Korea}

\author{Young-Kyun Kwon}
\email{ykkwon@khu.ac.kr}
\affiliation{Department of Physics and Research Institute for Basic Sciences, Kyung Hee University, Seoul 02447, Korea}
\affiliation{Department of Information Display, Kyung Hee University, Seoul 02447, Korea}

\begin{abstract}
Predicting the full spin Hall conductivity tensor of a magnetic crystal is difficult because crystallographic and magnetic symmetries do not constrain all parts of the response in the same way. Here we separate the spin Hall conductivity of altermagnets into time-reversal-even and time-reversal-odd channels and combine symmetry analysis with first-principles calculations for six representative compounds spanning distinct crystallographic classes. The even channel follows crystallographic selection rules consistent with its Berry-curvature Fermi-sea origin. The odd channel is set by magnetic symmetry and can be reshaped by antiunitary operations whose spatial parts originate from screw or glide symmetries. Momentum-resolved analysis further links the even response to spin-orbit-driven avoided crossings and the odd response to anisotropic near-Fermi-level spin-current response textures. By converting crystallographic and magnetic symmetry operations into linear constraints on the full 27-component SHC tensor, this framework provides a computational pre-screening route for identifying symmetry-allowed spin Hall channels before dense first-principles transport calculations. Extending the same analysis to 62 spin-split collinear antiferromagnets yields a magnetic-point-group tensor atlas, identifying design limits such as the complete 13/14 tensor partition in CuF$_2$ and the one-parameter time-odd tensor compression in MnSe.
\end{abstract}

\maketitle

\section{Introduction}\label{sec1}

Spin-charge interconversion underlies a wide range of spintronic functions, from spin-current generation to electrical control of magnetic order \cite{otani2017spin}. Among the available mechanisms, the spin Hall effect (SHE) remains one of the most useful because in its conventional geometry it generates a transverse spin current from an applied electric field and can be integrated into practical device geometries \cite{kato2004observation,SHE,manipatruni2019scalable,wang2024inverse}. The response is described by the spin Hall conductivity (SHC) tensor, $\sigma_{ij}^{k}$, and the tensorial character immediately makes the symmetry decisive: symmetry determines which components can exist, which are related, and which must vanish \cite{seemann2015symmetry}. That same sensitivity also makes it difficult to infer the full tensor before a complete calculation, even when the electronic structure is already reasonably understood \cite{macneill2017control,derunova2019giant}.

A useful step in the theory of spin Hall transport in magnetic materials was the explicit decomposition of the SHC into time-reversal-even ($\mathcal{T}$-even) and time-reversal-odd ($\mathcal{T}$-odd) parts \cite{SHCS_EQ1,SC_ncl_AFM2,SHC_EQ1,SHCS_EQ2,SHC_EQ2,SHCS_EQ3,SHCS_EQ4,USHC_ref,my}. Throughout this work, these labels denote the parity of the response coefficient under reversal of the complete ordered magnetic state $\mathcal{M}$, where $\mathcal{TM}$ is obtained by reversing all local magnetic moments while keeping the crystal structure fixed. Following symmetry-based classifications, we use ``SHC tensor'' for the full rank-three coefficient $\sigma_{ij}^{k}$ relating $J_i^k$ to $E_j$, including symmetry-allowed conventional, collinear, and longitudinal configurations~\cite{USHC}; this geometrical classification is independent of its $\mathcal{T}$ parity. The two channels are not just different ways to group the same response. They originate from distinct microscopic processes and obey different symmetry constraints. The $\mathcal{T}$-even contribution is dominated by interband coherence accumulated over occupied states and is therefore tied to Berry-curvature physics and Fermi-sea response \cite{xiao2010berry}; it can remain finite even in nonmagnetic systems \cite{USHC}. The $\mathcal{T}$-odd contribution, by contrast, is weighted by states near the Fermi level and is much more sensitive to magnetic order, spin polarization, and the symmetry of the ordered state. Once the response is resolved this way, one can ask a sharper question: which parts of the tensor are fixed by crystallography and which require the full magnetic symmetry group?

Altermagnets provide an especially clean setting in which to address that question. They combine collinear compensated magnetic order with momentum-dependent spin splitting, which places them outside the conventional ferromagnet-antiferromagnet dichotomy \cite{alter_0,alter_a1,alter_a2}. In these systems, operations that preserve a spin sublattice and those that exchange opposite-spin sublattices can play qualitatively different roles, often through nonsymmorphic elements such as screw rotations and glide mirrors. As a consequence, altermagnets can host spin-split bands even without spin-orbit coupling (SOC), unusual momentum-space spin textures, and unconventional spin-transport responses \cite{my}. The field has also moved quickly in the last two years. Recent work has broadened the altermagnetic landscape from symmetry-focused overviews to metallic room-temperature realizations and transport studies in both collinear and chiral settings \cite{jungwirth2026signatures,jiang2025metallic,farajollahpour2025berry,hu2025spinhallchiral}. This makes transferable symmetry rules increasingly useful, especially for computational screening.

Experimental evidence consistent with altermagnetic behavior has been reported in MnTe \cite{MnTe_exp,alter_SOC} and CrSb \cite{CrSb}, and more recent work continues to expand the set of candidate or confirmed platforms. What remains less settled is how to infer the full SHC tensor once both crystallographic and magnetic symmetry are active. General symmetry classifications of spin-conductivity tensors already exist \cite{USHC,USHC_ref,seemann2015symmetry}, and our recent study clarified how crystal and magnetic symmetry control unconventional SHC in selected altermagnets \cite{my}. The aim here is broader. Rather than focusing on a few materials with especially visible unconventional responses, we ask how the $\mathcal{T}$-even and $\mathcal{T}$-odd sectors are organized across representative altermagnetic symmetry classes.

That gap matters for two reasons. First, the allowed $\mathcal{T}$-even tensor is often inherited from crystallography, whereas the $\mathcal{T}$-odd tensor can be restructured by antiunitary magnetic operations. Second, materials sharing a crystallographic class do not necessarily share the same $\mathcal{T}$-odd tensor once magnetic sublattice symmetry is included. In other words, the full SHC tensor of an altermagnet can, in general, not be inferred from crystal symmetry alone.

In this work, we combine symmetry analysis and first-principles calculations to build such a framework. We consider six representative altermagnets--FeSb$_2$ (mmm) \cite{FeSb2}, MnO$_2$ and MnF$_2$ (4/mmm) \cite{MnO2,MnF2}, CuF$_2$ (2/m) \cite{CuF2}, FeF$_3$ ($\bar{3}$m) \cite{FeF3}, and MnSe (6mm) \cite{MnSe,bezzerga2025high}--chosen to span different crystallographic classes as well as representative nonsymmorphic and antiunitary symmetries. For each material, we determine the symmetry-allowed $\mathcal{T}$-even and $\mathcal{T}$-odd tensor components, identify symmetry-enforced relations among them, and connect those tensor structures to their momentum-resolved microscopic origins. To evaluate the transferability of the tensor rules beyond this representative set, we further apply the same magnetic-point-group analysis to the 62 spin-split collinear antiferromagnets listed by Guo et al.~\cite{spin_split}; the resulting canonical tensor patterns are provided in the Supplementary Note~S3. The resulting symmetry hierarchy of SHC in altermagnets shows that crystallographic symmetry governs the $\mathcal{T}$-even channel, whereas magnetic symmetry, especially antiunitary nonsymmorphic operations, governs the $\mathcal{T}$-odd channel. 

\section{Results}\label{sec2}

\subsection{Symmetry-resolved picture of SHC in altermagnets}\label{subsec3}

\begin{figure}[t]
  \centering
  \includegraphics[width=1.00\linewidth]{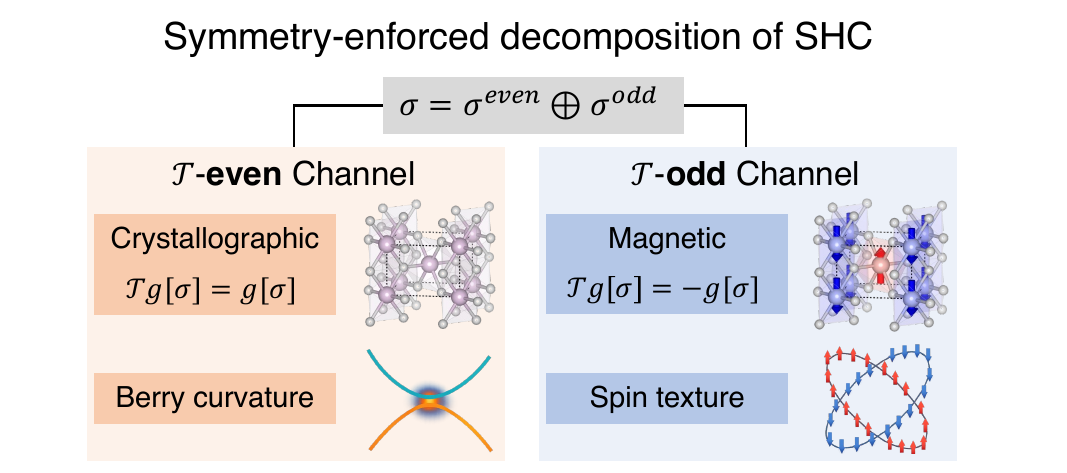}
  \caption{Symmetry-enforced decomposition of the spin Hall conductivity in altermagnets. The SHC tensor $\sigma_{ij}^{k}$ separates into $\mathcal{T}$-even and $\mathcal{T}$-odd channels governed by different symmetry constraints. In the $\mathcal{T}$-even channel, antiunitary operations of the form $\mathcal{T}g$ do not impose independent selection rules and reduce to the constraints of the corresponding unitary operation $g$. The allowed tensor components are therefore determined by crystallographic symmetry, and the response originates from Fermi-sea Berry curvature. In the $\mathcal{T}$-odd channel, by contrast, antiunitary operations provide genuinely additional constraints because the response changes sign under time reversal. This channel is therefore governed by magnetic symmetry and is microscopically tied to near-Fermi-level spin-current matrix elements and response textures.}
  \label{fig1}
\end{figure}

The starting point of the present analysis is that the SHC in altermagnets naturally separates into two channels with different symmetry content (Fig.~\ref{fig1}). For the $\mathcal{T}$-even part, the response is invariant under time reversal. In a magnetic crystal, the full tensor is in principle constrained by the magnetic point group, including antiunitary operations. However, once the response itself is even under $\mathcal{T}$, an operation of the form $\mathcal{T}g$ imposes the same selection rule as its unitary counterpart $g$. In practice, the allowed $\mathcal{T}$-even tensor components are therefore inherited from the parent crystallographic point group, just as in nonmagnetic systems. This is consistent with the microscopic origin of $\mathcal{T}$-even SHC as a Berry-curvature-driven Fermi-sea response.

The $\mathcal{T}$-odd channel behaves differently. Because it changes sign with time reversal, an antiunitary symmetry $\mathcal{T}g$ does not reduce to the corresponding unitary operation. It imposes an independent constraint and therefore the tensor form must be derived from the magnetic symmetry of the ordered state rather than from crystallography alone. This is precisely where altermagnets become interesting. Their symmetry operations often distinguish between preserving a spin sublattice and exchanging opposite-spin sublattices, and the latter may survive only as antiunitary symmetries once magnetic order is included. As a result, the $\mathcal{T}$-odd tensor can be suppressed, related, or activated in ways that are invisible at the crystallographic level. The practical consequence is straightforward: the $\mathcal{T}$-even tensor can be read from crystallographic selection rules, whereas the $\mathcal{T}$-odd tensor requires the full magnetic symmetry analysis. This symmetry separation also defines the computational workflow used in this study. Starting from the crystallographic structure and the assumed collinear magnetic configuration, we classify each spatial operation according to its action on the two magnetic sublattices and construct the corresponding unitary subgroup and antiunitary coset of the magnetic point group. Each operation is then represented as a \(27\times27\) matrix acting on the vectorized SHC tensor, and the allowed \(\mathcal{T}\)-even and \(\mathcal{T}\)-odd tensor spaces are obtained by solving the resulting linear constraints. Dense first-principles Brillouin-zone integrations are then performed for the symmetry-allowed tensor components and used to evaluate their magnitudes and microscopic momentum-space origins. In this way, the symmetry analysis acts as a pre-screening step for identifying target SHC channels in altermagnetic materials.

\begin{table}[t]
\centering
\caption{Classification of altermagnetic materials based on spin-momentum-locking classes and textures, together with the corresponding point groups and space groups. Representative materials considered in this work are listed for each class.}
\includegraphics[width=\linewidth]{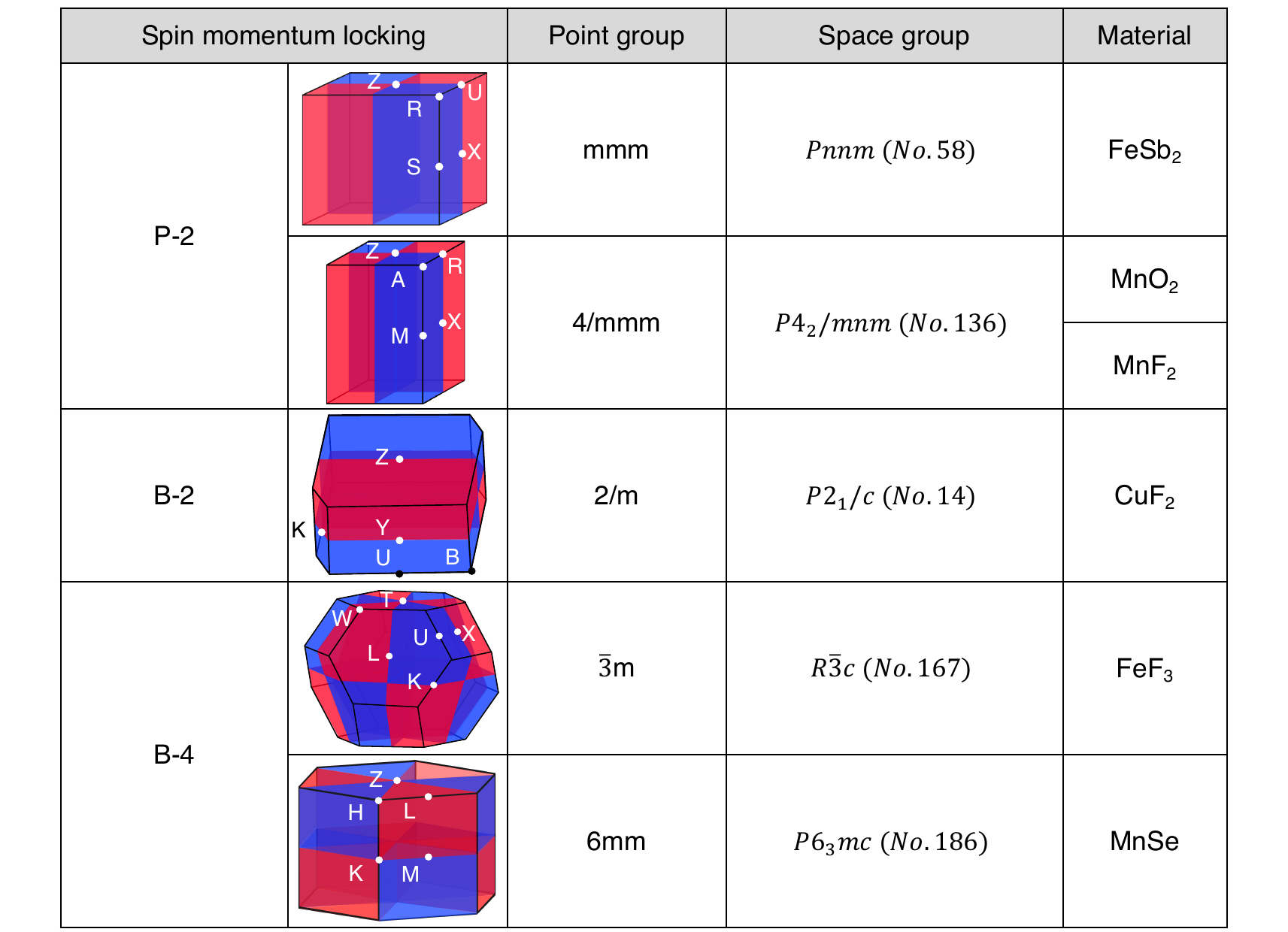}
\label{table1}
\end{table}

\subsection{Material set and symmetry classes}\label{subsec4}

To make this comparison as transparent as possible, we use a compact material set that spans the relevant crystallographic classes, spin-momentum-locking classes, and representative nonsymmorphic or antiunitary operations (Table~\ref{table1}; structural details in the Supplementary Note~S1). The set is not intended to be exhaustive. Instead, it is designed to isolate which tensor constraints are universal and which are material specific.

FeSb$_2$, MnO$_2$, and MnF$_2$ belong to class P-2 but cover orthorhombic and tetragonal crystallography. CuF$_2$ represents class B-2 with reduced monoclinic symmetry, where the lower crystallographic symmetry already enlarges the available tensor space. FeF$_3$ and MnSe belong to class B-4 with trigonal and hexagonal symmetry, respectively, but they differ markedly in the antiunitary and nonsymmorphic operations that survive in the ordered state. This combination provides a compact platform for disentangling what is controlled by crystallography from what is controlled by magnetic sublattice symmetry.

For \(\mathrm{FeF}_3\), this distinction is particularly relevant. In representative-material analysis, \(\mathrm{FeF}_3\) is used as a high-symmetry trigonal reference with the ordered moment constrained along the \(z\) axis, preserving the \(C_{3z}\) symmetry and the \(\bar{3}m\) point group. This reference configuration should not be confused with the experimentally reported in-plane Fe-moment structure, which lowers the ordered-state MPG to \(2^{\prime}/m^{\prime}\) and is treated separately in the Supplementary Note~S3.

\begin{table}[t]
\centering
\small
\setlength{\tabcolsep}{19pt}
\caption{Design-rule summary for symmetry-allowed SHC tensors in representative altermagnets. For each compound, we list the key symmetry operation and the number of symmetry-allowed tensor components for the $\mathcal{T}$-even and $\mathcal{T}$-odd channels ($n_{\mathrm{even}}$ and $n_{\mathrm{odd}}$). The first number gives the number of symmetry-allowed tensor components, and the parenthesized number gives the number of symmetry-independent parameters. Full component lists are given in Supplementary Note~S2. The \(\mathrm{FeF}_3\) entry is a high-symmetry trigonal reference with the ordered moment constrained along the \(z\) axis, preserving the \(C_{3z}\) symmetry and the \(\bar{3}m\) point group. The experimentally reported in-plane Fe-moment configuration lowers the ordered-state MPG to \(2^{\prime}/m^{\prime}\) and is analyzed separately in Supplementary Note~S3 under MPG $2'/m'$.}
\label{tab:designrule}
\begin{tabular}{@{}llll@{}}
\toprule
Material & Key symmetry operation & $n_{\mathrm{even}}$ allowed (ind.) & $n_{\mathrm{odd}}$ allowed (ind.) \\
\midrule
FeSb$_2$ & Orthorhombic symmetries & 6 (6) & 6 (6) \\
MnO$_2$  & $\mathcal{T}4_{2,z}$ & 6 (3) & 6 (3) \\
MnF$_2$  & $\mathcal{T}4_{2,z}$ & 6 (3) & 6 (3) \\
CuF$_2$  & $\mathcal{T}2_{1,z}$, $\mathcal{T}M_{z}$ & 13 (13) & 14 (14) \\
FeF$_3$  & $C_{3z}$ & 10 (4) & 10 (4) \\
MnSe     & $\mathcal{T}M_x$, $M_y(\tau_{z+1/2})$ & 6 (3) & 4 (1) \\
\bottomrule
\end{tabular}
\end{table}

Table~\ref{tab:designrule} already shows the central message. The $\mathcal{T}$-even and $\mathcal{T}$-odd sectors may share the same number of allowed components, as in FeSb$_2$, MnO$_2$, MnF$_2$ and FeF$_3$, yet the symmetry relations among these components need not be identical. In CuF$_2$, the two channels even differ in the number of allowed components, whereas in MnSe the $\mathcal{T}$-odd sector is strongly compressed by magnetic symmetry.

To examine the transferability of this symmetry hierarchy beyond the six representative compounds, we further applied the same magnetic-point-group analysis to the 62 spin-split collinear antiferromagnets listed by Guo~\textit{et al.}~\cite{spin_split}. This database-level analysis is summarized in the Supplementary Note~S3. Because several monoclinic and rhombohedral compounds in such databases may be reported in different crystallographic settings, all component labels in the survey are given in a canonical MPG coordinate convention with the ordered moment aligned along the canonical $z$ axis. We therefore organize the results in terms of canonical tensor patterns and allowed/independent component counts, rather than treating the listed Cartesian labels as material-specific crystallographic tensor components.

\subsection{Crystal-symmetry selection rules for the $\mathcal{T}$-even SHC}\label{subsec5}

\begin{figure}[t]
  \centering
  \includegraphics[width=1.00\linewidth]{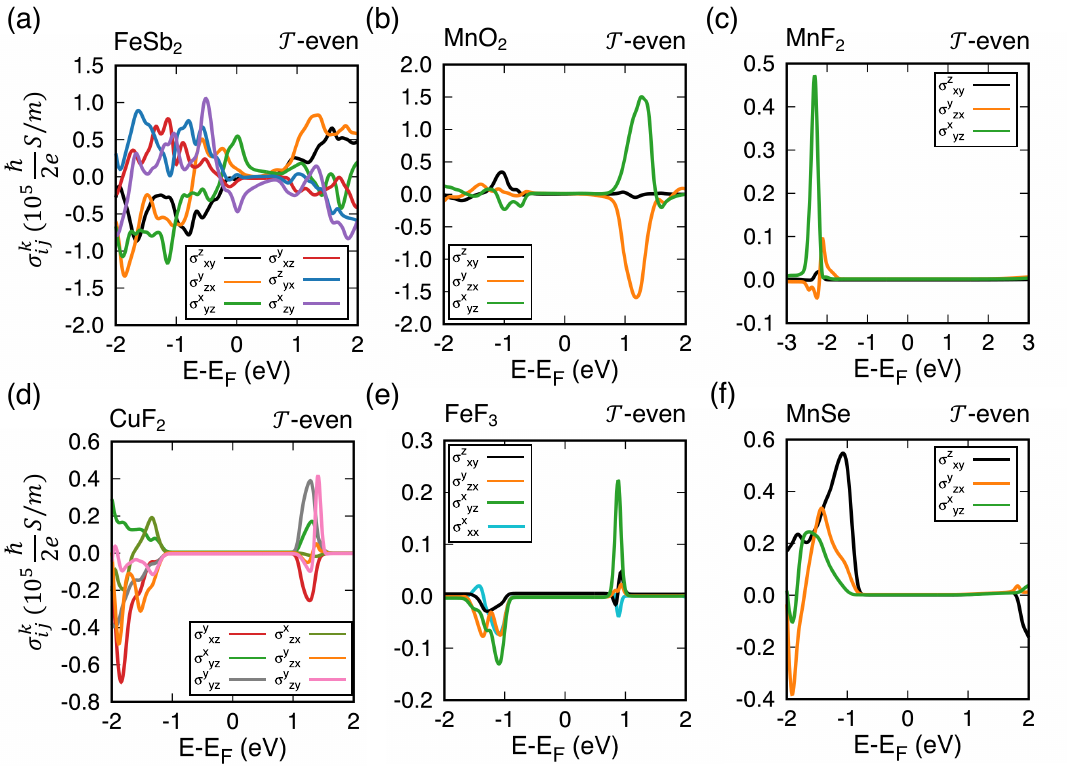}
  \caption{Energy dependence of the $\mathcal{T}$-even SHC components symmetry-allowed by crystal symmetry. Dominant components are highlighted where appropriate. (a) FeSb$_2$ (mmm): six symmetry-allowed and independent components. (b,c) MnO$_2$ and MnF$_2$ (4/mmm): six allowed components reduced to three independent ones by symmetry. (d) CuF$_2$ (2/m): thirteen allowed and independent components; only the dominant ones are shown. (e) FeF$_3$ ($\bar{3}$m): ten allowed components with four independent ones. (f) MnSe (6mm): six allowed components with three independent ones.}
  \label{fig2}
\end{figure}

Figure~\ref{fig2} summarizes the symmetry-allowed $\mathcal{T}$-even SHC tensor components for each point group. A clear trend emerges: in altermagnets, the $\mathcal{T}$-even SHC follows the same crystallographic selection rules as in nonmagnetic systems. This is the tensor-level manifestation of its Fermi-sea Berry-curvature origin.

FeSb$_2$ provides the simplest case. Because the orthorhombic point group lacks higher rotational symmetry, six conventional SHC components are allowed, and all six are independent. MnO$_2$ and MnF$_2$ share the tetragonal class 4/mmm, where six components remain allowed, but only three are independent because the fourfold screw-related symmetry enforces equivalence relations among in-plane tensor elements. FeF$_3$ shows an analogous reduction, now driven by $C_{3z}$, so that the ten allowed components collapse to four independent parameters. In MnSe, sixfold symmetry yields six allowed components and three independent ones.

CuF$_2$ is the most permissive case in the present set. Its monoclinic symmetry leaves thirteen $\mathcal{T}$-even tensor components symmetry-allowed and independent. This is exactly what one expects once higher rotational symmetry is removed: crystallography opens a larger tensor space, and the electronic structure then decides which among those allowed components become dominant.

This separation between ``which components may exist'' and ``how large they are'' is useful for design. For the $\mathcal{T}$-even channel, crystallography fixes the former, whereas the details of SOC-driven interband mixing fix the latter. Materials with the same crystallographic symmetry can therefore share the same allowed tensor form while still showing very different SHC magnitudes.

\subsection{Magnetic-symmetry selection rules for the $\mathcal{T}$-odd SHC}\label{subsec6}

\begin{figure}[t]
  \centering
  \includegraphics[width=1.00\linewidth]{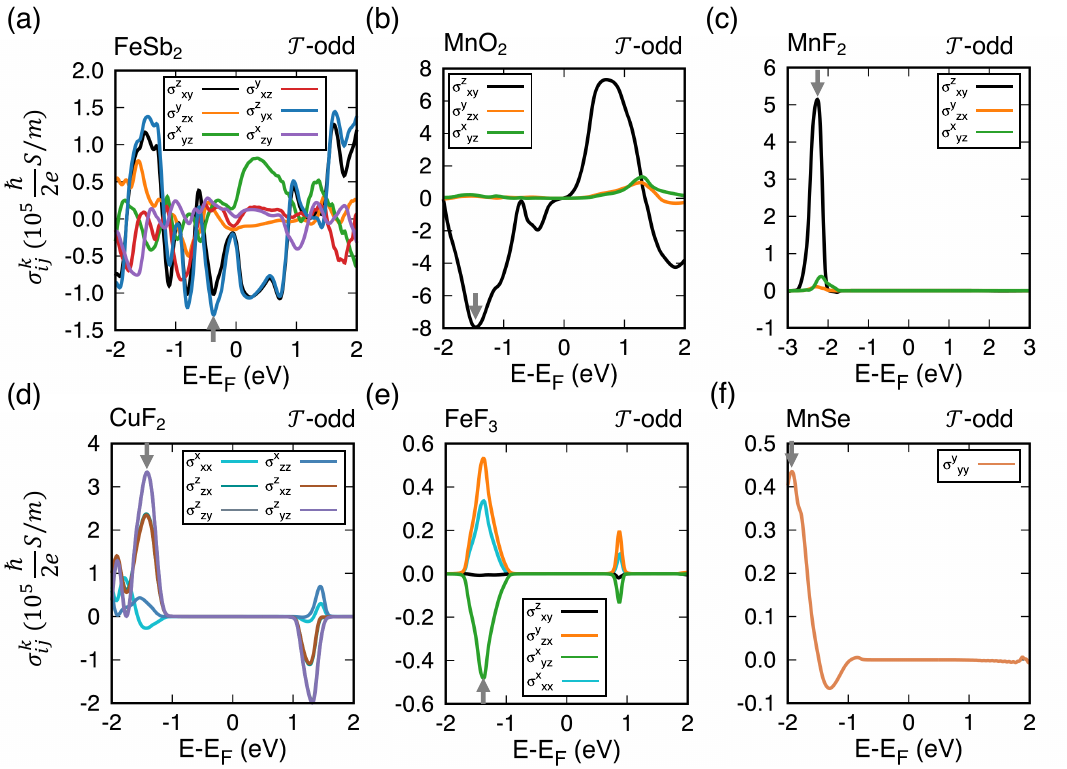}
  \caption{Energy dependence of the $\mathcal{T}$-odd SHC components symmetry-allowed by magnetic symmetry for moments aligned along $z$. Dominant components are highlighted where appropriate. (a) FeSb$_2$: six symmetry-allowed and independent components. (b,c) MnO$_2$ and MnF$_2$: six allowed components with three independent ones. (d) CuF$_2$: fourteen symmetry-allowed and independent components; only the dominant ones are shown. (e) FeF$_3$: ten allowed components with four independent ones. (f) MnSe: four allowed components with only one independent parameter. Gray arrows mark the energies used for the momentum-resolved $\mathcal{T}$-odd analysis in Fig.~\ref{fig5}.}
  \label{fig3}
\end{figure}

We next turn to the $\mathcal{T}$-odd contribution, whose tensor structure is governed by magnetic symmetry rather than crystallography (Fig.~\ref{fig3}). For a consistent comparison, we keep the ordered moments aligned along the crystallographic $z$ axis. Rotating the moment direction would modify the magnetic symmetry and therefore could change the allowed $\mathcal{T}$-odd tensor components, but it would not change the central logic of the analysis.

FeSb$_2$ again serves as a baseline. Introducing magnetic order does not generate additional antiunitary constraints beyond those already encoded in the crystallographic symmetry, and the $\mathcal{T}$-odd SHC therefore retains the same six allowed and independent components as the $\mathcal{T}$-even tensor.

MnO$_2$ and MnF$_2$ show the first important departure. In the magnetic state, the unitary screw rotation is not preserved as a standalone operation, but the combined antiunitary symmetry $\mathcal{T}4_{2,z}$ remains valid. This changes the tensor relations. The number of allowed components stays the same as in the $\mathcal{T}$-even sector, but the symmetry relations among the paired elements are modified because the antiunitary operation acts differently on a response that is odd under time reversal. In this sense, the tensor rank is unchanged, whereas the tensor algebra is not.

CuF$_2$ shows the strongest contrast between the two channels. Screw rotation $2_{1,z}$ and glide mirror $M_z(\tau_{x+1/2}+\tau_{y+1/2})$ survive only in antiunitary forms $\mathcal{T}2_{1,z}$ and $\mathcal{T}M_z$. As a result, some tensor elements that are forbidden in the $\mathcal{T}$-even sector become allowed in the $\mathcal{T}$-odd sector, while others behave in the opposite way. The effect is large enough that the two sectors no longer even contain the same number of allowed components: thirteen for $\mathcal{T}$-even and fourteen for $\mathcal{T}$-odd. More importantly, the two sets are exactly complementary: the thirteen $\mathcal{T}$-even and fourteen $\mathcal{T}$-odd components have no overlap and together exhaust all 27 entries of the rank-three SHC tensor. Components with even axial parity under $\mathcal{T}2_{1,z}$ and $\mathcal{T}M_z$---such as $\sigma^{x}_{yz}$, $\sigma^{x}_{zy}$, $\sigma^{y}_{xz}$, $\sigma^{y}_{zx}$, $\sigma^{z}_{xy}$, and $\sigma^{z}_{yx}$---populate the $\mathcal{T}$-even sector and reproduce the conventional SHE-type structure. The opposite-parity components---$\sigma^{x}_{xx}$, $\sigma^{x}_{zz}$, $\sigma^{y}_{yy}$, $\sigma^{z}_{xx}$, $\sigma^{z}_{yy}$, $\sigma^{z}_{zz}$, together with longitudinal pairs such as $(\sigma^{x}_{xz},\sigma^{x}_{zx})$ and $(\sigma^{y}_{xy},\sigma^{y}_{yx})$---are forbidden in $\mathcal{T}$-even but become symmetry-allowed in $\mathcal{T}$-odd. This complete partition of the tensor space is the cleanest tensor-level signature of antiunitary screw/glide-derived symmetry selection: although the translational parts of nonsymmorphic operations do not enter the homogeneous tensor transformation directly, they determine whether the associated spatial operation survives in the unitary subgroup or in the antiunitary coset. This is the clearest example in our set of how antiunitary nonsymmorphic operations can activate unconventional SHC channels.

FeF$_3$ returns to a more conventional case. Magnetic symmetry does not qualitatively alter the tensor structure in relation to crystallographic prediction, and the $\mathcal{T}$-odd channel allows the same ten components as the $\mathcal{T}$-even channel, with four independent parameters. MnSe represents the opposite limit. Once magnetic order is imposed, one mirror operation is replaced by its antiunitary counterpart, while the remaining symmetry elements strongly constrain the tensor. Only four $\mathcal{T}$-odd components survive and reduce to a single independent parameter.

Taken together, these material-resolved examples establish a simple design rule. The $\mathcal{T}$-odd SHC is a sensitive probe of magnetic symmetry in altermagnets, and antiunitary operations built from time reversal and nonsymmorphic elements can decisively determine which tensor elements survive. A complete list of symmetry-allowed $\mathcal{T}$-even and $\mathcal{T}$-odd components for six representative materials is provided in the Supplementary Note~S2.

\subsection{Microscopic origin of the $\mathcal{T}$-even response: SOC-driven interband mixing}\label{subsec7}

\begin{figure}[t]
  \centering
  \includegraphics[width=1.00\linewidth]{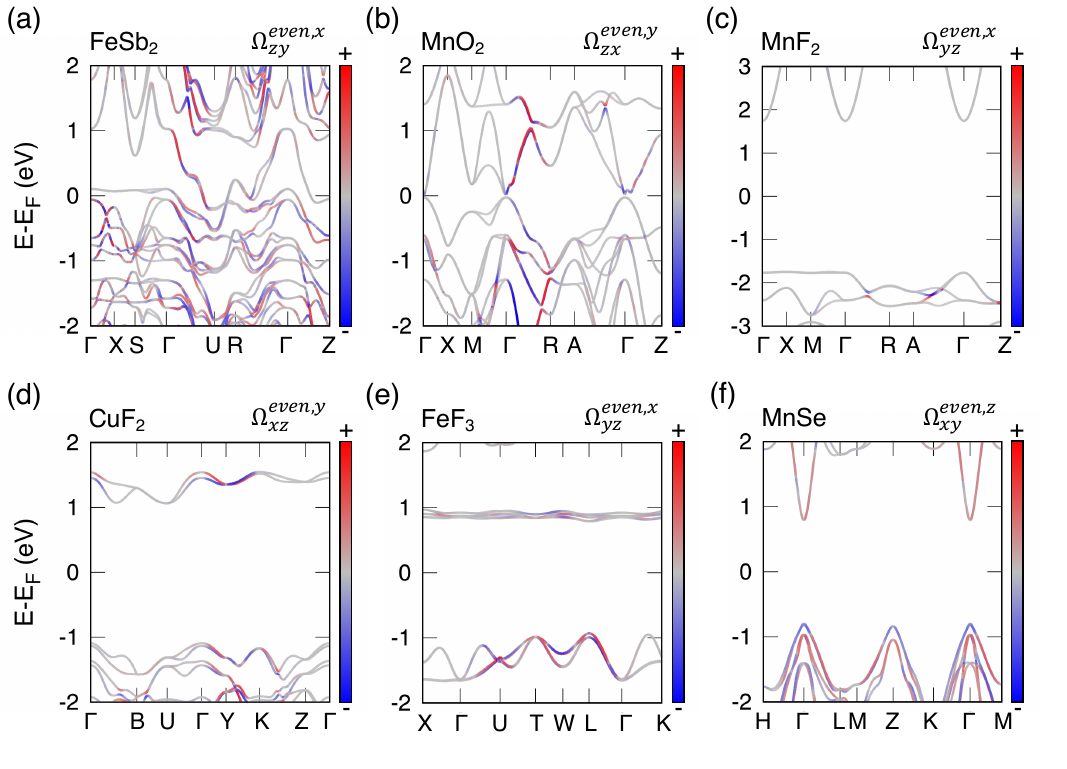}
  \caption{Band structures overlaid with the logarithmically scaled representative $\mathcal{T}$-even Berry-curvature-like term $\Omega^{\mathrm{even},k}_{ij}(\mathbf{k})$. (a) $\Omega^{\mathrm{even},x}_{zy}$ of FeSb$_2$, (b) $\Omega^{\mathrm{even},y}_{zx}$ of MnO$_2$, (c) $\Omega^{\mathrm{even},x}_{yz}$ of MnF$_2$, (d) $\Omega^{\mathrm{even},y}_{xz}$ of CuF$_2$, (e) $\Omega^{\mathrm{even},x}_{yz}$ of FeF$_3$, and (f) $\Omega^{\mathrm{even},z}_{xy}$ of MnSe. The strongest contributions occur near SOC-induced avoided crossings, underscoring the role of interband mixing in the $\mathcal{T}$-even SHC.}
  \label{fig4}
\end{figure}

To connect the tensor classification with microscopic mechanisms, we next inspect the momentum- and energy-resolved Berry-curvature-like quantities for each channel. Figure~\ref{fig4} overlays the representative $\Omega^{\mathrm{even},k}_{ij}$ on the band structures along the high-symmetry lines.

Across all compounds, the $\mathcal{T}$-even contribution is sharply localized in momentum and energy space. Its largest values appear near SOC-induced avoided crossings or in regions where several bands come close in energy. This is what one expects for an interband-coherence-driven response: the signal is amplified where SOC hybridization is strongest and energy denominators become small.

The detailed hotspot pattern is material dependent. In FeSb$_2$, the signal is distributed over several SOC-hybridized regions, producing a relatively broad pattern. Instead, MnO$_2$ and MnF$_2$ show more localized peaks associated with symmetry-line gap openings. CuF$_2$ exhibits a particularly sharp hotspot near Y, reflecting a SOC-lifted near-degeneracy. FeF$_3$ shows several intense features along extended high-symmetry paths, while MnSe presents a broader response that extends below the Fermi level.

The overall message is consistent with the tensor analysis in Fig.~\ref{fig2}. Crystal symmetry decides which $\mathcal{T}$-even tensor elements may be finite, but the actual magnitude is set by the electronic structure---most notably by where SOC produces strong interband mixing.

\subsection{Microscopic origin of the $\mathcal{T}$-odd response: near-Fermi-level spin-current matrix elements}\label{subsec8}

\begin{figure}[t]
  \centering
  \includegraphics[width=1.00\linewidth]{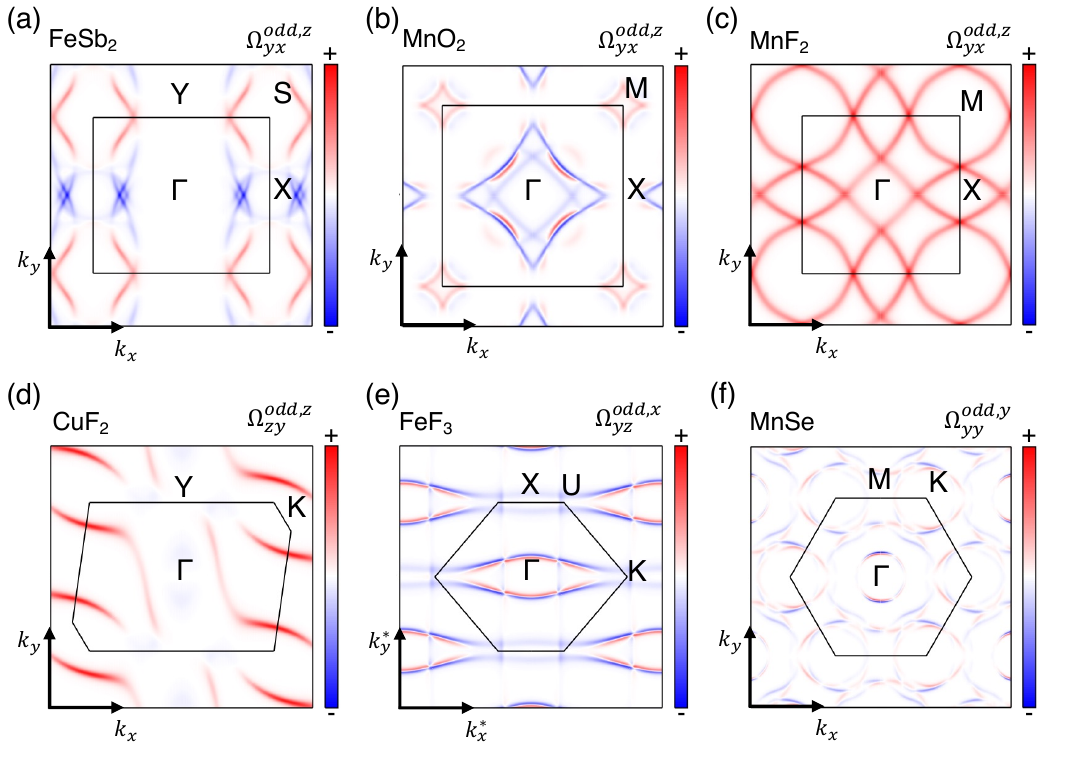}
  \caption{Momentum-resolved distribution of the representative $\mathcal{T}$-odd Berry-curvature-like term $\Omega^{\mathrm{odd},k}_{ij}(\mathbf{k})$. (a) $\Omega^{\mathrm{odd},z}_{yx}$ of FeSb$_2$, (b) $\Omega^{\mathrm{odd},z}_{yx}$ of MnO$_2$, (c) $\Omega^{\mathrm{odd},z}_{yx}$ of MnF$_2$, (d) $\Omega^{\mathrm{odd},z}_{yz}$ of CuF$_2$, (e) $\Omega^{\mathrm{odd},x}_{yz}$ of FeF$_3$, and (f) $\Omega^{\mathrm{odd},y}_{yy}$ of MnSe. In (e), $k_x^*$ and $k_y^*$ denote in-plane projections along reciprocal vectors $\mathbf{b}_1$ and $\mathbf{b}_2$. The patterns reveal highly anisotropic momentum-space structures that dominate the $\mathcal{T}$-odd response.}
  \label{fig5}
\end{figure}

The $\mathcal{T}$-odd channel looks qualitatively different. Figure~\ref{fig5} shows the momentum-resolved $\Omega^{\mathrm{odd},k}_{ij}(\mathbf{k})$ on representative two-dimensional Brillouin-zone cuts chosen from the energies marked in Fig.~\ref{fig3}. Instead of the localized avoided-crossing hotspots characteristic of the $\mathcal{T}$-even response, the $\mathcal{T}$-odd contribution exhibits pronounced sign-alternating structures, nodal lines, and strongly anisotropic lobes. These structures reflect the magnetic symmetry of the ordered state and the associated near-Fermi-level spin-current matrix elements.

FeSb$_2$, MnO$_2$, and MnF$_2$ all show structured patterns around $\Gamma$, but the detailed shapes vary from localized alternating patches to extended arc-like features. CuF$_2$ displays enhanced elongated diagonal structures near the zone boundaries, consistent with the strong role of antiunitary nonsymmorphic symmetry found at the tensor level. FeF$_3$ exhibits stripe-like features aligned with in-plane reciprocal directions, whereas MnSe shows a much weaker and more compressed pattern, consistent with the strong symmetry suppression of its $\mathcal{T}$-odd tensor.

These distributions make the microscopic origin of the $\mathcal{T}$-odd SHC visually explicit. The response is not generated by a smooth background over the Brillouin zone. It is shaped by highly anisotropic, symmetry-constrained near-Fermi-level spin textures and the corresponding spin-current matrix elements. Combined with the tensor analysis of Fig.~\ref{fig3}, this establishes a coherent picture: the unconventional SHC of altermagnets is fundamentally a magnetic-symmetry-driven response.

\section{Discussion}\label{sec3}

The six-material comparison makes the symmetry hierarchy concrete. The $\mathcal{T}$-even sector is inherited from crystallography and behaves much like the conventional intrinsic SHC of nonmagnetic crystals. The $\mathcal{T}$-odd sector is different in character and in practice. Its tensor form depends on which magnetic operations remain valid in the ordered state, especially when time reversal must be combined with screw or glide operations. That is why materials with the same crystallographic class can still show different $\mathcal{T}$-odd SHC tensors. The $\mathcal{T}$-odd tensor thus represents the part of the spin-conductivity response that reverses sign upon reversal of the complete ordered magnetic state and vanishes when pure time reversal is a symmetry. Although Ref.~\cite{SC_ncl_AFM2} reserved the term “spin Hall effect” for the $\mathcal{T}$-even spin current, subsequent literature has widely used “$\mathcal{T}$-odd spin Hall conductivity/effect” for the odd response \cite{RuO2_2_tilt,Cao2023,Dai2024,zhu2024}. Its transverse antisymmetric part is commonly referred to as the magnetic spin Hall effect~\cite{USHC_ref}.

This frame also clarifies how the present work differs from recent material-specific studies. Our earlier work on unconventional SHC in selected altermagnets \cite{my} focused on how crystal and magnetic symmetry shape prominent unconventional responses in a small set of compounds. Here, the goal is broader and, in a computational-material sense, more transferable: we explicitly separate the $\mathcal{T}$-even and $\mathcal{T}$-odd sectors and track how their tensor forms evolve across representative altermagnetic symmetry classes. Recent transport-oriented work has likewise broadened the altermagnetic context, from Berry-curvature-based probes of altermagnetic order to large spin Hall and Edelstein responses in chiral non-collinear systems \cite{farajollahpour2025berry,hu2025spinhallchiral}. The present framework sits naturally alongside those studies because it supplies the tensor-level selection rules needed before one asks how large a given response may become in a particular material.


The practical implication is simple. If one wants to screen a candidate altermagnet for conventional spin-charge conversion, the crystallographic class already tells which $\mathcal{T}$-even tensor channels are available. If the target is instead a magnetic-symmetry-controlled response, the screening must include ordered-state symmetry from the beginning. The database-level extension to the Guo material list demonstrates how this logic can be used as a practical prescreening tool. Within a chosen magnetic configuration and canonical MPG coordinate convention, the MPG determines whether a desired $\mathcal{T}$-even or $\mathcal{T}$-odd SHC tensor pattern is symmetry-allowed; dense first-principles transport calculations are then needed only for the surviving channels and for evaluating their magnitudes. In that sense, the workflow is computationally reusable: identifying the relevant crystallographic and magnetic operations, determining the allowed tensor sector, and then investing in dense Brillouin-zone calculations for the target components. The six case studies here provide the microscopic origin of the response, while the Guo-list extension shows how the same symmetry rules scale to a broader set of altermagnetic materials.

At the microscopic level, the same hierarchy persists. The $\mathcal{T}$-even channel is intensified by SOC-driven avoided crossings and interband mixing, whereas the $\mathcal{T}$-odd channel is assembled from anisotropic spin textures near the Fermi level. This separation suggests a concrete route for material design. Strong SOC and near-degenerate bands are the right ingredients to enhance the $\mathcal{T}$-even response, while the $\mathcal{T}$-odd response is better tuned by controlling the magnetic symmetry, moment direction, and the presence or absence of antiunitary nonsymmorphic operations. The framework therefore gives more than a classification. It provides a practical basis for screening altermagnets with targeted SHC tensors and for deciding which symmetry lever is worth tuning in a given material family.

\section{Methods}\label{sec4}

\subsection{First-principles calculations}\label{subsec0}

Density functional theory (DFT) calculations were performed using Quantum ESPRESSO \cite{hohenberg1964inhomogeneous,kohn1965self,QE-2009,QE-2017} with projector augmented-wave (PAW) pseudopotentials \cite{blochl1994projector} and the Perdew-Burke-Ernzerhof generalized-gradient approximation \cite{perdew1996generalized}. Throughout, a plane-wave kinetic-energy cutoff of 60~Ry was used throughout. Brillouin-zone integrations employed $16\times16\times16$ Monkhorst-Pack $k$-point meshes for FeSb$_2$, MnO$_2$, MnF$_2$, CuF$_2$, FeF$_3$, and MnSe. Spin-orbit coupling was included in all calculations. On-site Coulomb interactions were treated within DFT+$U$: $U=4$~eV for the Mn $3d$ states \cite{li2019intrinsic,chen2019topological,lai2021defect,MnTeU2} and $U=5$~eV, $J=1$~eV for the Cu $3d$ states \cite{alter_0}.

\subsection{Wannier interpolation and SHC evaluation}\label{subsec1}

To evaluate SHC in dense $k$ meshes, we construct spinor maximally localized Wannier functions \cite{marzari1997maximally,mostofi2014updated} and the corresponding Wannier tight-binding Hamiltonians. Initial projections included $sp^3d^2$ and $d_{xy}/d_{xz}/d_{yz}$-like orbitals for Fe, Mn, and Cu, and $p$ orbitals for O, Sb, F, and Se. The upper bound of the frozen window was set to approximately 6~eV above the Fermi level. SHC was evaluated using WannierBerri \cite{tsirkin2021high} on $100\times100\times100$ $k$-point grids with a lifetime broadening $\Gamma \approx 50$~meV. The results were cross-checked against an independent Wannier-based linear-response implementation \cite{SC_ncl_AFM2}. The numerical protocol follows our previous SHC workflow, where k-mesh, broadening, and Wannierization convergence were benchmarked systematically~\cite{my}.

\subsection{SHC formalism and symmetry analysis}\label{subsec2}

Within the Kubo formalism \cite{SHCS_EQ1,SC_ncl_AFM2,SHC_EQ1,SHCS_EQ2,SHC_EQ2,SHCS_EQ3,USHC_ref,SHCS_EQ4,my}, the SHC is decomposed into $\mathcal{T}$-even and $\mathcal{T}$-odd contributions,
\begin{equation}
\sigma_{ij}^{\mathrm{even},k} = -e\hbar \int_{\mathrm{BZ}}\frac{d^3k}{(2\pi)^3}
\sum_{n\neq m}(f_{n\mathbf{k}}-f_{m\mathbf{k}})
\frac{\mathrm{Im}\left(\langle n\mathbf{k}|\frac{1}{2}\{\hat{s}_k,\hat{v}_i\}|m\mathbf{k}\rangle
\langle m\mathbf{k}|\hat{v}_j|n\mathbf{k}\rangle\right)}{(\epsilon_{n\mathbf{k}}-\epsilon_{m\mathbf{k}})^2},
\label{eq:shc_even}
\end{equation}
\begin{equation}
\sigma_{ij}^{\mathrm{odd},k} = \frac{-e\hbar}{2\pi}\Gamma^2 \int_{\mathrm{BZ}}\frac{d^3k}{(2\pi)^3}
\sum_{n,m}
\frac{\mathrm{Re}\left(\langle n\mathbf{k}|\frac{1}{2}\{\hat{s}_k,\hat{v}_i\}|m\mathbf{k}\rangle
\langle m\mathbf{k}|\hat{v}_j|n\mathbf{k}\rangle\right)}{[(E_\mathrm{F}-\epsilon_{n\mathbf{k}})^2+\Gamma^2][(E_\mathrm{F}-\epsilon_{m\mathbf{k}})^2+\Gamma^2]},
\label{eq:shc_odd}
\end{equation}
where $\hat{s}^k$ and $\hat{v}_{i,j}$ denote the spin and velocity operators, respectively, and $\Gamma$ is the lifetime broadening parameter.



The superscripts ``even'' and ``odd'' denote the behavior of the response coefficient under reversal of the ordered magnetic state:
\begin{equation*}
\sigma_{ij}^{\pm,k}[\mathcal{M}]
=\frac{1}{2}\left[
\sigma_{ij}^{k}[\mathcal{M}]
\pm\sigma_{ij}^{k}[\mathcal{TM}]
\right].
\end{equation*}
Although the conventional spin-current operator $\hat{Q}_{i}^{k}=\tfrac{1}{2}\{\hat{s}_{k},\hat{v}_{i}\}$ is $\mathcal{T}$-even, the velocity vertex is $\mathcal{T}$-odd and time reversal is antiunitary. Denoting the matrix-element product in the two Kubo terms by $C_{nm}^{kij}(\mathbf{k};\mathcal{M})$, one obtains
\begin{equation*}
C_{\bar n\bar m}^{kij}(-\mathbf{k};\mathcal{TM})
=-\left[C_{nm}^{kij}(\mathbf{k};\mathcal{M})\right]^{*}.
\end{equation*}
Its real and imaginary parts therefore generate the $\mathcal{T}$-odd and $\mathcal{T}$-even contributions, respectively, consistent with the parity assignment for spin conductivity in Ref.~\cite{SC_ncl_AFM2}.

Equations~\eqref{eq:shc_even} and \eqref{eq:shc_odd} define the $\mathcal{T}$-even and $\mathcal{T}$-odd contributions, respectively. Within the present Kubo formulation, their sum gives the SHC response considered in this work.

The $\mathcal{T}$-even contribution originates from interband coherence among occupied states and has the character of a Fermi-sea response. It can therefore remain finite even in nonmagnetic systems. For the collinear altermagnetic configurations considered here, the selection rules for this channel coincide with those obtained from the crystallographic point group, because an antiunitary operation of the form $\mathcal{T}g$ acts on a $\mathcal{T}$-even response in the same way as its unitary part $g$. By contrast, the $\mathcal{T}$-odd contribution has a Fermi-surface-like character and is concentrated near the Fermi level. It is therefore sensitive to the symmetry of the ordered magnetic state. Because the two channels have opposite time-reversal parity, antiunitary magnetic symmetry operations impose different constraints on them~\cite{USHC_ref}.

Under a point-group operation $g$ represented by a $3\times3$ orthogonal matrix $D(g)$ with $\det D(g)=\pm1$, the SHC tensor transforms as,
\begin{equation}
\sigma_{ij}^{k}\mapsto \det D(g)\,D(g)_{ii'}\,D(g)_{kk'}\,D(g)_{jj'}\,\sigma_{i'j'}^{k'},
\end{equation}
where the factor $\det D(g)$ reflects the axial-vector character of the spin index. For a homogeneous transport tensor, the translation part of a nonsymmorphic operation does not enter this tensor transformation directly. Instead, its role is to determine whether the corresponding spatial operation belongs to the unitary subgroup or to the antiunitary coset of the magnetic point group in the ordered magnetic structure. 

Imposing the transformation rule of Eq.~(3) on $\sigma_{ij}^{k}$ yields the symmetry-allowed tensor components~\cite{seemann2015symmetry}. We implement this as a linear-constraint problem in the full 27-dimensional tensor space. The components of $\sigma_{ij}^{k}$ are collected in a vector $\vec{\sigma}$, and each spatial operation $g$ is represented by the induced $27\times27$ matrix $\rho_{\sigma}(g)$ obtained from Eq.~(3); equivalently, $\rho_{\sigma}(g)$ is the representation induced from the $3\times3$ matrix $D(g)$ on the rank-three tensor space.

Writing the magnetic point group as $M = H \cup \mathcal{T}\cdot A$, the unitary subgroup $H$ and the antiunitary coset $\mathcal{T}A$ impose different constraints depending on the time-reversal parity of the response. For $g\in H$,
\begin{equation}
\rho_{\sigma}(g)\vec{\sigma}=\vec{\sigma},
\end{equation}
whereas for $g\in A$,
\begin{equation}
\rho_{\sigma}(g)\vec{\sigma}=\pm\vec{\sigma},
\end{equation}
with the upper (lower) sign for the $\mathcal{T}$-even ($\mathcal{T}$-odd) channel. Thus, antiunitary operations reduce to their unitary spatial constraints for the $\mathcal{T}$-even tensor, but impose opposite-sign constraints for the $\mathcal{T}$-odd tensor.

For an altermagnet with collinear moments along $\hat{\bm{z}}$, each parent operation is assigned to $H$ or $\mathcal{T}A$ according to its combined action on $\sigma_z$ and on the two magnetic sublattices. The resulting linear system is solved by singular-value decomposition and validated against the nonmagnetic SHC classification of Roy~\textit{et al.}~\cite{USHC} for the $\mathcal{T}$-even sector, with the redundant antiunitary constraints omitted. A formal derivation, the sublattice classification table, and the validation benchmark are provided in the Supplementary Note~S2. 

For the database-level extension to the Guo altermagnet list, we used the MPG labels and magnetic space-group types reported ~\cite{spin_split}. The unitary subgroup $H$ and the antiunitary coset $\mathcal{T}A$ are constructed according to the canonical coordinate convention of the MPG, with the ordered moment taken to define the canonical axis $z$. The resulting tensor forms are reported as canonical component patterns rather than material-specific Cartesian tensors. The canonical patterns and the mapping of all 62 compounds on the Guo-list are provided in the Supplementary Note~S3.

In the SHC tensor $\sigma_{ij}^{k}$, the indices $j$, $i$, and $k$ denote the directions of the applied electric field, the spin-current flow, and the spin polarization, respectively. To clarify the microscopic origin of each channel, we also analyze momentum-resolved Berry-curvature-like quantities,
\begin{equation}
\Omega^{\mathrm{even},k}_{ij}(\mathbf{k})=-\sum_{n\neq m}
\frac{\mathrm{Im}\left[
\langle n|\frac{1}{2}\{\hat{s}^k,\hat{v}_i\}|m\rangle
\langle m|\hat{v}_j|n\rangle\right]}{(\epsilon_n-\epsilon_m)^2},
\label{eq:Omega_even}
\end{equation}
\begin{equation}
\Omega^{\mathrm{odd},k}_{ij}(\mathbf{k})=-\Gamma^2\sum_{n,m}
\frac{\mathrm{Re}\left[\langle n|\frac{1}{2}\{\hat{s}^k,\hat{v}_i\}|m\rangle\langle m|\hat{v}_j|n\rangle\right]}{[(E_F-\epsilon_n)^2+\Gamma^2][(E_F-\epsilon_m)^2+\Gamma^2]}.
\label{eq:Omega_odd}
\end{equation}
Equation~\eqref{eq:Omega_even} captures the $\mathcal{T}$-even Berry-curvature-like contribution, whereas Eq.~\eqref{eq:Omega_odd} captures the $\mathcal{T}$-odd contribution arising from states near the Fermi level.

\begin{acknowledgments}
This work was supported by the National Research Foundation of Korea (NRF) grant funded by the Korean government (MSIT) (RS-2022-NR069250 and RS-2024-00416976). S.-H. Kang was partially supported by the Korea Institute of Science and Technology Information (KISTI) (K26L4M2C2-01). Part of the computational work was carried out using the resources of the KISTI Supercomputing Center (KSC-2024-CRE-0540). 

\end{acknowledgments}

\section*{Data Availability}
The processed data used to generate the figures, including the energy-dependent SHC curves, momentum-resolved response maps, and symmetry-allowed tensor tables, will be made publicly available in a suitable data repository upon publication. Additional raw first-principles data is available from the corresponding authors upon reasonable request.

\section*{Code Availability}
The scripts used for the linear-constraint symmetry analysis of the SHC tensor will be made publicly available in a suitable code repository upon publication. Quantum ESPRESSO, Wannier90, and WannierBerri are publicly available. Additional custom scripts used for post-processing and figure generation are available from the corresponding authors upon reasonable request.

\section*{Author Contributions}
D.J., S.-H.K., and Y.-K.K. conceived the project. D.J. carried out the first-principles calculations and SHC analysis. S.-H.K. contributed to the symmetry analysis and interpretation. All authors discussed the results and contributed to the writing of the manuscript. S.-H.K. and Y.-K.K. supervised the project.

\section*{Competing Interests}
The authors declare no competing interests.

\clearpage

\setcounter{section}{0}
\setcounter{subsection}{0}
\setcounter{subsubsection}{0}
\setcounter{equation}{0}
\setcounter{figure}{0}
\setcounter{table}{0}

\renewcommand{\theequation}{S\arabic{equation}}
\renewcommand{\thefigure}{S\arabic{figure}}
\renewcommand{\thetable}{S\arabic{table}}
\renewcommand{\thesection}{S\arabic{section}}
\renewcommand{\thesubsection}{S\arabic{section}.\Alph{subsection}}
\renewcommand{\thesubsubsection}{S\arabic{section}.\Alph{subsection}.\arabic{subsubsection}}

\makeatletter
\renewcommand{\p@subsection}{}
\renewcommand{\p@subsubsection}{}
\renewcommand{\p@paragraph}{}
\renewcommand{\p@subparagraph}{}
\makeatother

\renewcommand{\theHequation}{S\arabic{equation}}
\renewcommand{\theHfigure}{S\arabic{figure}}
\renewcommand{\theHtable}{S\arabic{table}}
\renewcommand{\theHsection}{S\arabic{section}}
\renewcommand{\theHsubsection}{S\arabic{section}.\Alph{subsection}}
\renewcommand{\theHsubsubsection}{S\arabic{section}.\Alph{subsection}.\arabic{subsubsection}}

\begin{center}
{\large\bfseries Supplemental Material for}\\[0.5em]
{\large\bfseries ``Computational symmetry hierarchy of time-reversal-even and -odd spin Hall conductivity tensors in altermagnets''}\\[1em]
Dameul Jeong, Seoung-Hun Kang, and Young-Kyun Kwon
\end{center}

\vspace{1em}
\section{Crystal Structure}\label{S1}

\begin{figure}[htb!]
\centering
\includegraphics[width=\columnwidth]{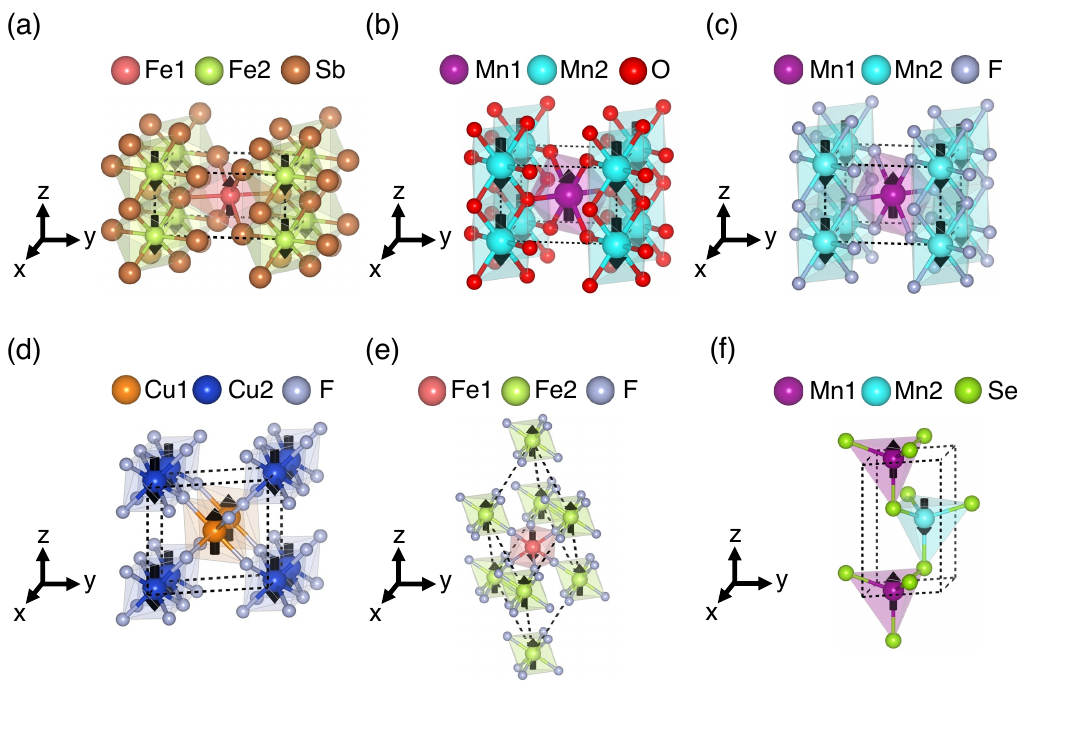}
\caption{Crystal structures of representative altermagnetic materials considered in this work. (a) FeSb$_2$, (b) MnO$_2$, (c) MnF$_2$, (d) CuF$_2$, (e) FeF$_3$, and (f) MnSe. Distinct magnetic sublattices are indicated by different colors for the transition-metal ions, while anions are shown in lighter tones. The dashed black lines denote the conventional crystallographic unit cells, and the coordinate axes are shown for reference. The corresponding Laue groups are FeSb$_2$ (mmm), MnO$_2$ (4/mmm), MnF$_2$ (4/mmm), CuF$_2$ (2/m), FeF$_3$ ($\bar{3}$m), and MnSe (6mm).}
\label{FIGS1}
\end{figure}
The crystallographic structures and lattice parameters of all materials investigated in this work are summarized below and visualized in Fig.~\ref{FIGS1}. Fig.~\ref{FIGS1}(a) shows the orthorhombic crystal structure of FeSb$_2$, characterized by lattice constants $a_1 = 5.82$~\AA, $a_2 = 6.57$~\AA, and $a_3 = 3.25$~\AA. Figs.~\ref{FIGS1}(b) and \ref{FIGS1}(c) display the tetragonal structures of MnO$_2$ and MnF$_2$, respectively, with lattice constants $a_1 = a_2 = 4.50$~\AA~and $a_3 = 2.95$~\AA~for MnO$_2$, and $a_1 = a_2 = 5.15$~\AA~and $a_3 = 3.38$~\AA~for MnF$_2$.

Fig.~\ref{FIGS1}(d) presents the monoclinic crystal structure of CuF$_2$, with lattice constants $a_1 = 3.41$~\AA, $a_2 = 5.37$~\AA, and $a_3 = 4.62$~\AA, where the angle between $a_1$ and $a_2$ is $120.69^\circ$. The rhombohedral structure of FeF$_3$ is shown in Fig.~\ref{FIGS1}(e), characterized by a lattice constant $a = 5.45$~\AA~and a rhombohedral angle of $58.64^\circ$. Finally, Fig.~\ref{FIGS1}(f) shows the hexagonal structure of MnSe, with lattice constants $a = 4.28$~\AA~and $a_3 = 6.87$~\AA.

In Fig.~\ref{FIGS1}, conventional crystallographic unit cells are indicated by dashed black lines, and the crystallographic axes are shown for reference. Distinct transition-metal sites that form different magnetic sublattices are highlighted by different colors, while ligand atoms are depicted with smaller spheres. This figure provides a direct real-space visualization of the lattice symmetries underlying the symmetry analysis presented in the main text and clarifies how the different crystal systems—orthorhombic, tetragonal, monoclinic, rhombohedral, and hexagonal—determine the associated Laue-group symmetries and the symmetry-allowed tensor components discussed in this work.
\clearpage
\section{Allowed SHC tensor Components}\label{S2}

\subsection{Symmetry-analysis framework}\label{S2:framework}
  
The spin Hall conductivity (SHC) tensor $\sigma^{k}_{ij}$ is a rank-three response tensor that relates the spin current $J_i^k$ to the applied electric field $E_j$ through $J_i^k=\sigma^k_{ij}E_j$. Although $J_i^k$ and $E_j$ are both even under time reversal, this bare-operator statement does not fix the magnetic-state parity of $\sigma_{ij}^{k}[\mathcal{M}]$, where $\mathcal{M}$ denotes the ordered magnetic state; throughout this work, that parity is defined by comparing the response coefficients of $\mathcal{M}$ and $\mathcal{TM}$, as detailed in the Methods. Here, the index $i$ denotes the flow direction of the spin current and the index $j$ denotes the electric-field direction. Both are polar-vector indices, because the velocity/current direction and the electric field transform in the same way as an ordinary displacement vector under spatial operations. By contrast, the spin-polarization index $k$ is an axial-vector index, because spin transforms as an angular momentum. Equivalently, spin has the same transformation character as $\bm{r}\times\bm{p}$, the cross product of two polar vectors.

Under a point-group operation $g$, we denote by $D(g)$ the
corresponding $3\times3$ orthogonal representation matrix acting on
the polar vectors in Cartesian coordinates, $v_i \mapsto D(g)_{ii'}\,v_{i'}$.
Thus, $\det D(g)=+1$ for proper rotations and $\det D(g)=-1$
for improper operations such as mirrors, inversion, and rotoinversions.
While a polar vector transforms only with $D(g)$, an axial vector
transforms with an additional determinant factor,
$S_k \mapsto \det D(g)\,D(g)_{kk'}\,S_{k'}$.
Combining the two polar indices $i,j$ and the one axial spin index $k$,
the SHC tensor transforms as
\begin{equation}\label{eq:sigma_transform}
\sigma^{k}_{ij}\;\longmapsto\;\det D(g)\,D(g)_{ii'}\,D(g)_{kk'}\,D(g)_{jj'}\,
\sigma^{k'}_{i'j'},
\end{equation}
with summation over the primed indices implied. The factor
$\det D(g)$ therefore accounts for the axial character of the spin
index, whereas the polar indices $i$ and $j$ transform without any
additional determinant factor.
 
For a magnetic crystal, the magnetic point group can be written as
\(M=H\cup \mathcal{T}A\), where \(H\) is the unitary subgroup and
\(\mathcal{T}A\) is the antiunitary coset when antiunitary operations are
present. Here, \(A\) denotes the set of spatial operations \(g\) that
appear in the antiunitary operations \(\mathcal{T}g\). For type-I groups,
\(A=\varnothing\) and \(H=M\). Following the \(\mathcal{T}\)-parity
decomposition of the spin conductivity tensor into Fermi-sea
(\(\mathcal{T}\)-even) and Fermi-surface (\(\mathcal{T}\)-odd)
contributions, a tensor component is symmetry-allowed only if the full magnetic operation acting on it leaves it invariant.

For \(g\in H\), the full magnetic operation is the spatial operation \(g\)
itself. For \(g\in A\), the full magnetic operation is the antiunitary
operation \(\mathcal{T}g\), not \(g\) alone. The spatial part \(g\)
transforms the tensor indices according to Eq.~(\ref{eq:sigma_transform}),
whereas \(\mathcal{T}\) contributes only the time-reversal parity of the
response. Therefore, the \(\mathcal{T}\)-even contribution satisfies
\begin{subequations}\label{eq:sea_rule}
\begin{align}
g:\quad
\sigma^{\mathrm{sea},k}_{ij}
&\mapsto
\bigl(g\!\cdot\!\sigma^{\mathrm{sea}}\bigr)^k_{ij}
=
\sigma^{\mathrm{sea},k}_{ij},
& g&\in H,\\[2pt]
\mathcal{T}g:\quad
\sigma^{\mathrm{sea},k}_{ij}
&\mapsto
+\,\bigl(g\!\cdot\!\sigma^{\mathrm{sea}}\bigr)^k_{ij}
=
\sigma^{\mathrm{sea},k}_{ij},
& g&\in A,
\end{align}
\end{subequations}
whereas the \(\mathcal{T}\)-odd contribution satisfies
\begin{subequations}\label{eq:surf_rule}
\begin{align}
g:\quad
\sigma^{\mathrm{surf},k}_{ij}
&\mapsto
\bigl(g\!\cdot\!\sigma^{\mathrm{surf}}\bigr)^k_{ij}
=
\sigma^{\mathrm{surf},k}_{ij},
& g&\in H,\\[2pt]
\mathcal{T}g:\quad
\sigma^{\mathrm{surf},k}_{ij}
&\mapsto
-\,\bigl(g\!\cdot\!\sigma^{\mathrm{surf}}\bigr)^k_{ij}
=
\sigma^{\mathrm{surf},k}_{ij},
& g&\in A.
\end{align}
\end{subequations}
Here, \((g\!\cdot\!\sigma)\) denotes only the spatial tensor
transformation defined in Eq.~(\ref{eq:sigma_transform}). The final equality
to \(\sigma\) in Eqs.~(\ref{eq:sea_rule})--(\ref{eq:surf_rule}) is the
invariance condition required for the tensor component to survive the
magnetic symmetry. The minus sign in the \(A\) coset for
\(\sigma^{\mathrm{surf}}\) reflects the odd time-reversal parity of the
Fermi-surface contribution. Thus, in the antiunitary coset, a
\(\mathcal{T}\)-odd tensor component can survive only when the spatial part
\(g\) compensates the sign change from time reversal. These constraints are
imposed on the full rank-three tensor and generate the complete set of
symmetry-allowed components, including their symmetric pieces and the
relations among them.

\subsection{Sublattice classification for altermagnets}\label{S2:sublattice}
 
For an altermagnetic ground state with two collinear magnetic sublattices
$A$ ($\uparrow$) and $B$ ($\downarrow$) aligned along $\hat{\bm{z}}$,
each parent operation $g$ falls into one of the four cases listed in
Table~\ref{tab:sublattice}, determined by its action on $\sigma_{z}$
and on the sublattice labels.
 
\begin{table}[h]
\centering
\renewcommand{\arraystretch}{1.1}
\caption{Assignment of a parent operation $g$ to the unitary subgroup
$H$ or the antiunitary coset $\mathcal{T}\!\cdot\!A$ of the magnetic
point group of an altermagnet with $\bm{M}\parallel\pm\hat{\bm{z}}$.}
\label{tab:sublattice}
\begin{tabular}{ccc}
\hline\hline
Action on $\sigma_{z}$ & Action on sublattice & MPG assignment\\
\hline
preserve & preserve $(A\!\to\!A,\,B\!\to\!B)$ & $H$ (unitary)\\
preserve & exchange $(A\!\leftrightarrow\!B)$ & $\mathcal{T}\!\cdot\!A$ (antiunitary)\\
flip     & preserve                           & $\mathcal{T}\!\cdot\!A$ (antiunitary)\\
flip     & exchange                           & $H$ (unitary)\\
\hline\hline
\end{tabular}
\end{table}
 
 
\subsection{Numerical implementation and validation}
\label{S2:implementation}

For numerical symmetry analysis, we first rewrite the SHC tensor as a
finite-dimensional vector. Before imposing symmetry constraints, the
rank-three tensor \(\sigma^{k}_{ij}\) has \(3\times3\times3=27\)
components. We gather these components into a single vector
\(\vec{\sigma}\) using the linear index
\[
\ell=9k+3i+j,
\]
where \(i,j,k=0,1,2\) corresponds to \(x,y,z\), respectively. With this
convention, the nine components with fixed spin polarization \(k=x\)
come first, followed by those with \(k=y\) and \(k=z\).

Each symmetry operation \(g\) is then represented by its induced
\(27\times27\) matrix \(\rho_{\sigma}(g)\) acting on the vectorized SHC
tensor. The entries of \(\rho_{\sigma}(g)\) are obtained directly from
the tensor transformation rule in Eq.~(\ref{eq:sigma_transform}). Thus,
\(\rho_{\sigma}(g)\) simply encodes how the operation \(g\) permutes,
relates, or changes the signs of the 27 tensor components.

A symmetry-allowed tensor must satisfy all symmetric constraints of the
magnetic point group. For a unitary operation \(g\in H\), the tensor must
remain unchanged after applying \(g\), which gives
\[
\vec{\sigma}=\rho_{\sigma}(g)\vec{\sigma}.
\]
For an antiunitary operation \(\mathcal{T}g\) with \(g\in A\), the
constraint depends on the time-reversal parity of the response. The
\(\mathcal{T}\)-even channel is invariant under time reversal, whereas
the \(\mathcal{T}\)-odd channel changes sign. Therefore, the allowed
tensor space is obtained from the linear constraints
\begin{subequations}
\label{eq:numerical_constraints}
\begin{align}
\bigl[\rho_{\sigma}(g)-\mathbf{I}_{27}\bigr]\vec{\sigma}&=0,
& g&\in H, \\
\bigl[\rho_{\sigma}(g)\mp\mathbf{I}_{27}\bigr]\vec{\sigma}&=0,
& g&\in A.
\label{eq:A}
\end{align}
\end{subequations}
Here, the upper sign applies to the \(\mathcal{T}\)-even channel and the
lower sign applies to the \(\mathcal{T}\)-odd channel in Eq.~(\ref{eq:A}). The matrix \(\mathbf{I}_{27}\) denotes the
\(27\times27\) identity matrix in the vector space of the SHC tensor components.

The resulting linear constraints are solved numerically using
singular-value decomposition. This procedure determines which tensor
components are forced to vanish, which surviving components are related
by symmetry, and how many independent tensor parameters remain.

To validate the implementation, we consider the nonmagnetic gray magnetic group \(G+\mathcal{T}G\), for which \(H=G\) and \(A=G\). For the \(\mathcal{T}\)-even SHC sector, an antiunitary operation \(\mathcal{T}g\) imposes the same tensor constraint as its unitary spatial part \(g\). The redundant antiunitary constraints are therefore omitted in this benchmark, so that the analysis reduces to the ordinary crystallographic selection rules. The resulting tensor forms agree with the exhaustive nonmagnetic classification of Roy~\textit{et al.}~\cite{USHC}, as summarized in Table~\ref{tab:validation}. In the full gray magnetic group, pure time reversal additionally forces the \(\mathcal{T}\)-odd SHC sector to vanish identically. Thus, Table~S2 benchmarks the \(\mathcal{T}\)-even crystallographic selection rules implemented in the present framework.

\begin{table}[h]
\centering
\renewcommand{\arraystretch}{1.1}
\caption{Benchmark of the present framework against the nonmagnetic SHC
classification of Roy~\textit{et al.}~\cite{USHC} for selected Laue
classes. Entries are given as allowed components / linearly independent
components.}
\label{tab:validation}
\begin{tabular}{lcc}
\hline\hline
Laue class & Roy~\textit{et al.}~\cite{USHC} & This work\\
\hline
$\bar{3}$    & 21\,/\,9 & 21\,/\,9\\
$\bar{3}m$   & 10\,/\,4 & 10\,/\,4\\
$4/mmm$      &  6\,/\,3 &  6\,/\,3\\
$6/mmm$      &  6\,/\,3 &  6\,/\,3\\
\hline\hline
\end{tabular}
\end{table}


\subsection{Application to representative altermagnetic materials}\label{S2:materials}
 
Applying Eqs.~(\ref{eq:sigma_transform})--(\ref{eq:surf_rule}) together
with the sublattice classification of Table~\ref{tab:sublattice} to the
magnetic point groups of the six altermagnets considered here, we obtain
the symmetry-allowed $\sigma^{\mathrm{sea}}$ and $\sigma^{\mathrm{surf}}$
tensors summarized in the following pages. For each material the
structure of the allowed-component matrices, the number of independent
parameters, and the explicit linear relations among components are
reported.

\begin{table}[h]
\centering
\setlength{\tabcolsep}{6pt}

\begin{tabular}{c c}

\begin{minipage}{0.48\textwidth}
\centering
\rule{\linewidth}{0.5pt}

\vspace{4pt}
\textbf{FeSb$_2$}\\
$\mathcal{T}$-even

\vspace{8pt}

\scriptsize
\renewcommand{\arraystretch}{0.9}

\begin{tabular}{c@{\hspace{1pt}}c@{\hspace{1pt}}c}

$\overset{\sigma^{x}}{
\begin{pmatrix}
0&0&0\\
0&0&\sigma^x_{yz}\\
0&\sigma^x_{zy}&0
\end{pmatrix}}$ &

$\overset{\sigma^{y}}{
\begin{pmatrix}
0&0&\sigma^y_{xz}\\
0&0&0\\
\sigma^y_{zx}&0&0
\end{pmatrix}}$ &

$\overset{\sigma^{z}}{
\begin{pmatrix}
0&\sigma^z_{xy}&0\\
\sigma^z_{yx}&0&0\\
0&0&0
\end{pmatrix}}$

\end{tabular}

\normalsize
\vspace{6pt}

6 components, 6 independent

\vspace{-30pt}
\[
\begin{aligned}
\end{aligned}
\]

\vspace{-20pt}
\rule{\linewidth}{0.5pt}
\end{minipage}

&
\begin{minipage}{0.48\textwidth}
\centering
\rule{\linewidth}{0.5pt}

\vspace{4pt}
\textbf{FeSb$_2$}\\
$\mathcal{T}$-odd

\vspace{8pt}

\scriptsize
\renewcommand{\arraystretch}{0.9}

\begin{tabular}{c@{\hspace{1pt}}c@{\hspace{1pt}}c}

$\overset{\sigma^{x}}{
\begin{pmatrix}
0&0&0\\
0&0&\sigma^x_{yz}\\
0&\sigma^x_{zy}&0
\end{pmatrix}}$ &

$\overset{\sigma^{y}}{
\begin{pmatrix}
0&0&\sigma^y_{xz}\\
0&0&0\\
\sigma^y_{zx}&0&0
\end{pmatrix}}$ &

$\overset{\sigma^{z}}{
\begin{pmatrix}
0&\sigma^z_{xy}&0\\
\sigma^z_{yx}&0&0\\
0&0&0
\end{pmatrix}}$

\end{tabular}

\normalsize
\vspace{6pt}

6 components, 6 independent

\vspace{-30pt}
\[
\begin{aligned}
\end{aligned}
\]

\vspace{-20pt}
\rule{\linewidth}{0.5pt}
\end{minipage}

\end{tabular}
\end{table}

\begin{table}[h]
\centering
\setlength{\tabcolsep}{6pt}

\begin{tabular}{c c}

\begin{minipage}{0.48\textwidth}
\centering
\rule{\linewidth}{0.5pt}

\vspace{4pt}
\textbf{MnO$_2$ \& MnF$_2$}\\
$\mathcal{T}$-even

\vspace{8pt}

\scriptsize
\renewcommand{\arraystretch}{0.9}

\begin{tabular}{c@{\hspace{1pt}}c@{\hspace{1pt}}c}

$\overset{\sigma^{x}}{
\begin{pmatrix}
0&0&0\\
0&0&\sigma^x_{yz}\\
0&\sigma^x_{zy}&0
\end{pmatrix}}$ &

$\overset{\sigma^{y}}{
\begin{pmatrix}
0&0&\sigma^y_{xz}\\
0&0&0\\
\sigma^y_{zx}&0&0
\end{pmatrix}}$ &

$\overset{\sigma^{z}}{
\begin{pmatrix}
0&\sigma^z_{xy}&0\\
\sigma^z_{yx}&0&0\\
0&0&0
\end{pmatrix}}$

\end{tabular}

\normalsize
\vspace{6pt}

6 components, 3 independent

\vspace{-18pt}
\[
\begin{aligned}
\sigma^z_{xy}=-\sigma^z_{yx}, \sigma^y_{zx}=-\sigma^x_{zy}, \sigma^x_{yz}=-\sigma^y_{xz}\\ 
\end{aligned}
\]

\vspace{-12pt}
\rule{\linewidth}{0.5pt}
\end{minipage}

&
\begin{minipage}{0.48\textwidth}
\centering
\rule{\linewidth}{0.5pt}

\vspace{4pt}
\textbf{MnO$_2$ \& MnF$_2$}\\
$\mathcal{T}$-odd

\vspace{8pt}

\scriptsize
\renewcommand{\arraystretch}{0.9}

\begin{tabular}{c@{\hspace{1pt}}c@{\hspace{1pt}}c}

$\overset{\sigma^{x}}{
\begin{pmatrix}
0&0&0\\
0&0&\sigma^x_{yz}\\
0&\sigma^x_{zy}&0
\end{pmatrix}}$ &

$\overset{\sigma^{y}}{
\begin{pmatrix}
0&0&\sigma^y_{xz}\\
0&0&0\\
\sigma^y_{zx}&0&0
\end{pmatrix}}$ &

$\overset{\sigma^{z}}{
\begin{pmatrix}
0&\sigma^z_{xy}&0\\
\sigma^z_{yx}&0&0\\
0&0&0
\end{pmatrix}}$

\end{tabular}

\normalsize
\vspace{6pt}

6 components, 3 independent

\vspace{-18pt}
\[
\begin{aligned}
\sigma^z_{xy}=\sigma^z_{yx}, \sigma^y_{zx}=\sigma^x_{zy}, \sigma^x_{yz}=\sigma^y_{xz}\\ 
\end{aligned}
\]

\vspace{-12pt}
\rule{\linewidth}{0.5pt}
\end{minipage}

\end{tabular}
\end{table}


\begin{table}[h]
\centering
\setlength{\tabcolsep}{6pt}

\begin{tabular}{c c}

\begin{minipage}{0.48\textwidth}
\centering
\rule{\linewidth}{0.5pt}

\vspace{4pt}
\textbf{CuF$_2$}\\
$\mathcal{T}$-even

\vspace{8pt}

\scriptsize
\renewcommand{\arraystretch}{0.9}

\begin{tabular}{c@{\hspace{-3pt}}c@{\hspace{-3pt}}c}

$\overset{\sigma^{x}}{
\begin{pmatrix}
0 & \sigma^x_{xy} & 0 \\
\sigma^x_{yx} & 0 & \sigma^x_{yz} \\
0 & \sigma^x_{zy} & 0
\end{pmatrix}}$ &

$\overset{\sigma^{y}}{
\begin{pmatrix}
\sigma^y_{xx} & 0 & \sigma^y_{xz} \\
0 & \sigma^y_{yy} & 0 \\
\sigma^y_{zx} & 0 & \sigma^y_{zz}
\end{pmatrix}}$ &

$\overset{\sigma^{z}}{
\begin{pmatrix}
0 & \sigma^z_{xy} & 0 \\
\sigma^z_{yx} & 0 & \sigma^z_{yz} \\
0 & \sigma^z_{zy} & 0
\end{pmatrix}}$ 

\end{tabular}

\normalsize
\vspace{6pt}

13 components, 13 independent

\vspace{-30pt}
\[
\begin{aligned}
\end{aligned}
\]

\vspace{-20pt}
\rule{\linewidth}{0.5pt}
\end{minipage}

&
\begin{minipage}{0.48\textwidth}
\centering
\rule{\linewidth}{0.5pt}

\vspace{4pt}
\textbf{CuF$_2$}\\
$\mathcal{T}$-odd

\vspace{8pt}

\scriptsize
\renewcommand{\arraystretch}{0.9}

\begin{tabular}{c@{\hspace{-3pt}}c@{\hspace{-3pt}}c}

$\overset{\sigma^{x}}{
\begin{pmatrix}
\sigma^x_{xx} & 0 & \sigma^x_{xz} \\
0 & \sigma^x_{yy} & 0 \\
\sigma^x_{zx} & 0 & \sigma^x_{zz}
\end{pmatrix}}$&

$\overset{\sigma^{y}}{
\begin{pmatrix}
0 & \sigma^y_{xy} & 0 \\
\sigma^y_{yx} & 0 & \sigma^y_{yz} \\
0 & \sigma^y_{zy} & 0
\end{pmatrix}}$ &

$\overset{\sigma^{z}}{
\begin{pmatrix}
\sigma^z_{xx} & 0 & \sigma^z_{xz} \\
0 & \sigma^z_{yy} & 0 \\
\sigma^z_{zx} & 0 & \sigma^z_{zz}
\end{pmatrix}}$

\end{tabular}

\normalsize
\vspace{6pt}

14 components, 14 independent

\vspace{-30pt}
\[
\begin{aligned}
\end{aligned}
\]

\vspace{-20pt}
\rule{\linewidth}{0.5pt}
\end{minipage}

\end{tabular}
\end{table}
\clearpage

\begin{table}[h]
\centering
\setlength{\tabcolsep}{6pt}

\begin{tabular}{c c}

\begin{minipage}{0.48\textwidth}
\centering
\rule{\linewidth}{0.5pt}

\vspace{4pt}
\textbf{FeF$_3$}\\
$\mathcal{T}$-even

\vspace{8pt}

\scriptsize
\renewcommand{\arraystretch}{0.9}

\begin{tabular}{c@{\hspace{-3pt}}c@{\hspace{-3pt}}c}

$\overset{\sigma^{x}}{
\begin{pmatrix}
\sigma^x_{xx} & 0 & 0 \\
0 & \sigma^x_{yy} & \sigma^x_{yz} \\
0 & \sigma^x_{zy} & 0
\end{pmatrix}}$ &

$\overset{\sigma^{y}}{
\begin{pmatrix}
0 & \sigma^y_{xy} & \sigma^y_{xz} \\
\sigma^y_{yx} & 0 & 0 \\
\sigma^y_{zx} & 0 & 0
\end{pmatrix}}$ &

$\overset{\sigma^{z}}{
\begin{pmatrix}
0 & \sigma^z_{xy} & 0 \\
\sigma^z_{yx} & 0 & 0 \\
0 & 0 & 0
\end{pmatrix}}$

\end{tabular}

\normalsize
\vspace{6pt}

10 components, 4 independent

\vspace{-18pt}
\[
\begin{aligned}
\sigma^x_{xx}=-\sigma^x_{yy}=-\sigma^y_{yx}=-\sigma^y_{xy},\\
\sigma^z_{xy}=-\sigma^z_{yx}, \sigma^y_{zx}=-\sigma^x_{zy}, \sigma^x_{yz}=-\sigma^y_{xz}\\ 
\end{aligned}
\]

\vspace{-6pt}
\rule{\linewidth}{0.5pt}
\end{minipage}

&
\begin{minipage}{0.48\textwidth}
\centering
\rule{\linewidth}{0.5pt}

\vspace{4pt}
\textbf{FeF$_3$}\\
$\mathcal{T}$-odd

\vspace{8pt}

\scriptsize
\renewcommand{\arraystretch}{0.9}

\begin{tabular}{c@{\hspace{-3pt}}c@{\hspace{-3pt}}c}

$\overset{\sigma^{x}}{
\begin{pmatrix}
\sigma^x_{xx} & 0 & 0 \\
0 & \sigma^x_{yy} & \sigma^x_{yz} \\
0 & \sigma^x_{zy} & 0
\end{pmatrix}}$ &

$\overset{\sigma^{y}}{
\begin{pmatrix}
0 & \sigma^y_{xy} & \sigma^y_{xz} \\
\sigma^y_{yx} & 0 & 0 \\
\sigma^y_{zx} & 0 & 0
\end{pmatrix}}$ &

$\overset{\sigma^{z}}{
\begin{pmatrix}
0 & \sigma^z_{xy} & 0 \\
\sigma^z_{yx} & 0 & 0 \\
0 & 0 & 0
\end{pmatrix}}$

\end{tabular}

\normalsize
\vspace{6pt}

10 components, 4 independent

\vspace{-18pt}
\[
\begin{aligned}
\sigma^x_{xx}=-\sigma^x_{yy}=-\sigma^y_{yx}=-\sigma^y_{xy},\\
\sigma^z_{xy}=-\sigma^z_{yx}, \sigma^y_{zx}=-\sigma^x_{zy}, \sigma^x_{yz}=-\sigma^y_{xz}\\ 
\end{aligned}
\]

\vspace{-6pt}
\rule{\linewidth}{0.5pt}
\end{minipage}

\end{tabular}
\end{table}

\begin{table}[h]
\centering
\setlength{\tabcolsep}{6pt}

\begin{tabular}{c c}

\begin{minipage}{0.48\textwidth}
\centering
\rule{\linewidth}{0.5pt}

\vspace{4pt}
\textbf{MnSe}\\
$\mathcal{T}$-even

\vspace{8pt}

\scriptsize
\renewcommand{\arraystretch}{0.9}

\begin{tabular}{c@{\hspace{1pt}}c@{\hspace{1pt}}c}

$\overset{\sigma^{x}}{
\begin{pmatrix}
0&0&0\\
0&0&\sigma^x_{yz}\\
0&\sigma^x_{zy}&0
\end{pmatrix}}$ &

$\overset{\sigma^{y}}{
\begin{pmatrix}
0&0&\sigma^y_{xz}\\
0&0&0\\
\sigma^y_{zx}&0&0
\end{pmatrix}}$ &

$\overset{\sigma^{z}}{
\begin{pmatrix}
0&\sigma^z_{xy}&0\\
\sigma^z_{yx}&0&0\\
0&0&0
\end{pmatrix}}$

\end{tabular}

\normalsize
\vspace{6pt}

6 components, 3 independent

\vspace{-18pt}
\[
\begin{aligned}
\sigma^z_{xy}=-\sigma^z_{yx}, \sigma^y_{zx}=-\sigma^x_{zy}, \sigma^x_{yz}=-\sigma^y_{xz}\\ 
\end{aligned}
\]

\vspace{-12pt}
\rule{\linewidth}{0.5pt}
\end{minipage}

&
\begin{minipage}{0.48\textwidth}
\centering
\rule{\linewidth}{0.5pt}

\vspace{4pt}
\textbf{MnSe}\\
$\mathcal{T}$-odd

\vspace{8pt}

\scriptsize
\renewcommand{\arraystretch}{0.9}

\begin{tabular}{c@{\hspace{1pt}}c@{\hspace{1pt}}c}

$\overset{\sigma^{x}}{
\begin{pmatrix}
0&\sigma^x_{xy}&0\\
\sigma^x_{yx}&0&0\\
0&0&0
\end{pmatrix}}$ &

$\overset{\sigma^{y}}{
\begin{pmatrix}
\sigma^y_{xx}&0&0\\
0&\sigma^y_{yy}&0\\
0&0&0
\end{pmatrix}}$ &

$\overset{\sigma^{z}}{
\begin{pmatrix}
0&0&0\\
0&0&0\\
0&0&0
\end{pmatrix}}$

\end{tabular}

\normalsize
\vspace{6pt}

4 components, 1 independent

\vspace{-18pt}
\[
\begin{aligned}
\sigma^y_{yy}=-\sigma^x_{xy}=-\sigma^x_{yx}=-\sigma^y_{xx}\\
\end{aligned}
\]

\vspace{-12pt}
\rule{\linewidth}{0.5pt}
\end{minipage}

\end{tabular}
\end{table}
\clearpage
\section{Database-level extension to the altermagnet list}\label{S3}

To examine the transferability of the symmetry hierarchy discussed in the main text, we extend the same magnetic-point-group (MPG) analysis to the spin-split collinear antiferromagnets listed by Guo~\textit{et al.}~\cite{spin_split}. Their material list was obtained from experimentally reported magnetic structures in the MAGNDATA database and contains collinear antiferromagnets that allow momentum-dependent spin splitting in the absence of spin-orbit coupling. Here, we did not recompute the electronic structure or the magnitude of the spin Hall conductivity (SHC) for these compounds. Instead, we used the reported MPGs and magnetic space-group types as symmetry input and determined the allowed $\mathcal{T}$-even and $\mathcal{T}$-odd SHC tensor sectors.

All component labels in this section are given in a canonical MPG coordinate convention with the ordered moment aligned along the canonical $z$ axis, $\bm{M}\parallel\hat{\bm{z}}$. The listed tensor components $\sigma^k_{ij}$ therefore refer to this canonical MPG-axis convention. They should not be interpreted as material-specific Cartesian labels before mapping the canonical axes onto the structural setting used for a given compound. This distinction is especially important for monoclinic and rhombohedral systems, where different crystallographic settings may relabel the Cartesian axes. Such relabeling changes the names of the tensor components but not the number of allowed components or the number of independent tensor parameters.

FeF$_3$ illustrates this distinction. In the representative-material analysis of the main text, FeF$_3$ is treated as a high-symmetry trigonal reference in which the ordered moment preserves the $C_{3z}$ axis. In the Guo-list mapping, FeF$_3$ instead follows the reported in-plane Fe-moment configuration, which lowers the ordered-state MPG to $2'/m'$. Thus, the two FeF$_3$ entries correspond to different moment configurations, with the Guo-list tensor expressed in the canonical MPG frame.

For each MPG, we construct the unitary subgroup $H$ and the antiunitary coset $\mathcal{T}\!\cdot\!A$ following the same convention as in Sect.~\ref{S2}. For type-I magnetic space groups, the antiunitary coset is empty, and all MPG operations are treated as unitary. For type-III magnetic space groups, the unprimed operations define $H$, while the primed operations are assigned to $A$ and appear combined with time reversal. We then impose the linear constraints on the entire 27-dimensional tensor space of $\sigma^k_{ij}$. The resulting tensor forms are summarized in the following as canonical patterns. Because many compounds share the same MPG and several MPGs generate identical tensor patterns, the component tables are given once at the pattern level and then assigned to the material list.

\subsection{Canonical tensor patterns}

Table~\ref{tab:S3patterns1} and Table~\ref{tab:S3patterns2} define the canonical tensor patterns that appear in the Guo list. Each pattern contains the allowed components and the symmetry relations among them. When no relation is shown, all listed components are independent.

\begin{table}[p]
\centering
\scriptsize
\renewcommand{\arraystretch}{1.15}
\setlength{\tabcolsep}{4pt}
\caption{Canonical tensor patterns appearing in the database-level MPG
analysis. Component labels are given in the canonical MPG coordinate convention
with $\bm{M}\parallel\hat{\bm{z}}$.}
\label{tab:S3patterns1}
\begin{tabular}{c c p{0.56\textwidth} p{0.24\textwidth}}
\hline\hline
Pattern & Count & Allowed components & Relations \\
\hline

$\mathcal{P}_{1}$ & 27/27 &
All components $\sigma^k_{ij}$ with $i,j,k=x,y,z$. &
None \\

$\mathcal{P}_{2}$ & 6/6 &
$\sigma^x_{yz}$, $\sigma^x_{zy}$,
$\sigma^y_{xz}$, $\sigma^y_{zx}$,
$\sigma^z_{xy}$, $\sigma^z_{yx}$ &
None \\

$\mathcal{P}_{3}$ & 13/13 &
$\sigma^x_{xy}$, $\sigma^x_{yx}$,
$\sigma^x_{yz}$, $\sigma^x_{zy}$,
$\sigma^y_{xx}$, $\sigma^y_{xz}$,
$\sigma^y_{yy}$, $\sigma^y_{zx}$,
$\sigma^y_{zz}$,
$\sigma^z_{xy}$, $\sigma^z_{yx}$,
$\sigma^z_{yz}$, $\sigma^z_{zy}$ &
None \\

$\mathcal{P}_{4}$ & 14/14 &
$\sigma^x_{xx}$, $\sigma^x_{xz}$,
$\sigma^x_{yy}$, $\sigma^x_{zx}$,
$\sigma^x_{zz}$,
$\sigma^y_{xy}$, $\sigma^y_{yx}$,
$\sigma^y_{yz}$, $\sigma^y_{zy}$,
$\sigma^z_{xx}$, $\sigma^z_{xz}$,
$\sigma^z_{yy}$, $\sigma^z_{zx}$,
$\sigma^z_{zz}$ &
None \\

$\mathcal{P}_{5}$ & 7/7 &
$\sigma^x_{xz}$, $\sigma^x_{zx}$,
$\sigma^y_{yz}$, $\sigma^y_{zy}$,
$\sigma^z_{xx}$, $\sigma^z_{yy}$,
$\sigma^z_{zz}$ &
None \\

$\mathcal{P}_{6}$ & 7/7 &
$\sigma^x_{xy}$, $\sigma^x_{yx}$,
$\sigma^y_{xx}$, $\sigma^y_{yy}$,
$\sigma^y_{zz}$,
$\sigma^z_{yz}$, $\sigma^z_{zy}$ &
None \\

\hline\hline
\end{tabular}
\end{table}

\begin{table}[p]
\centering
\scriptsize
\renewcommand{\arraystretch}{1.15}
\setlength{\tabcolsep}{4pt}
\caption{Continuation of the canonical tensor patterns. The component labels
refer to the same canonical MPG coordinate convention as in
Table~\ref{tab:S3patterns1}.}
\label{tab:S3patterns2}
\begin{tabular}{c c p{0.47\textwidth} p{0.34\textwidth}}
\hline\hline
Pattern & Count & Allowed components & Relations \\
\hline

$\mathcal{P}_{7}$ & 6/3 &
$\sigma^x_{yz}$, $\sigma^x_{zy}$,
$\sigma^y_{xz}$, $\sigma^y_{zx}$,
$\sigma^z_{xy}$, $\sigma^z_{yx}$ &
$\sigma^x_{yz}=-\sigma^y_{xz}$,\newline
$\sigma^x_{zy}=-\sigma^y_{zx}$,\newline
$\sigma^z_{xy}=-\sigma^z_{yx}$ \\

$\mathcal{P}_{8}$ & 6/3 &
$\sigma^x_{yz}$, $\sigma^x_{zy}$,
$\sigma^y_{xz}$, $\sigma^y_{zx}$,
$\sigma^z_{xy}$, $\sigma^z_{yx}$ &
$\sigma^x_{yz}=+\sigma^y_{xz}$,\newline
$\sigma^x_{zy}=+\sigma^y_{zx}$,\newline
$\sigma^z_{xy}=+\sigma^z_{yx}$ \\

$\mathcal{P}_{9}$ & 4/1 &
$\sigma^x_{xy}$, $\sigma^x_{yx}$,
$\sigma^y_{xx}$, $\sigma^y_{yy}$ &
$\sigma^x_{xy}=-\sigma^y_{yy}$,\newline
$\sigma^x_{yx}=-\sigma^y_{yy}$,\newline
$\sigma^y_{xx}=-\sigma^y_{yy}$ \\

$\mathcal{P}_{10}$ & 10/4 &
$\sigma^x_{xx}$, $\sigma^x_{yy}$,
$\sigma^x_{yz}$, $\sigma^x_{zy}$,
$\sigma^y_{xy}$, $\sigma^y_{xz}$,
$\sigma^y_{yx}$, $\sigma^y_{zx}$,
$\sigma^z_{xy}$, $\sigma^z_{yx}$ &
$\sigma^x_{xx}=-\sigma^y_{yx}$,\newline
$\sigma^x_{yy}=+\sigma^y_{yx}$,\newline
$\sigma^x_{yz}=-\sigma^y_{xz}$,\newline
$\sigma^x_{zy}=-\sigma^y_{zx}$,\newline
$\sigma^y_{xy}=+\sigma^y_{yx}$,\newline
$\sigma^z_{xy}=-\sigma^z_{yx}$ \\

$\mathcal{P}_{11}$ & 10/4 &
$\sigma^x_{xy}$, $\sigma^x_{yx}$,
$\sigma^x_{yz}$, $\sigma^x_{zy}$,
$\sigma^y_{xx}$, $\sigma^y_{xz}$,
$\sigma^y_{yy}$, $\sigma^y_{zx}$,
$\sigma^z_{xy}$, $\sigma^z_{yx}$ &
$\sigma^x_{xy}=-\sigma^y_{yy}$,\newline
$\sigma^x_{yx}=-\sigma^y_{yy}$,\newline
$\sigma^x_{yz}=-\sigma^y_{xz}$,\newline
$\sigma^x_{zy}=-\sigma^y_{zx}$,\newline
$\sigma^y_{xx}=-\sigma^y_{yy}$,\newline
$\sigma^z_{xy}=-\sigma^z_{yx}$ \\

\hline\hline
\end{tabular}
\end{table}
\clearpage

\subsection{MPG-level assignment}

Table~\ref{tab:S3mpg} lists the unitary subgroup $H$ and the antiunitary coset $\mathcal{T}\!\cdot\!A$ used for each distinct MPG in the Guo list, together with the resulting canonical $\mathcal{T}$-even and $\mathcal{T}$-odd tensor patterns. The columns $|H|$ and $|A|$ give the number of unitary and antiunitary MPG elements obtained by closing these generators under composition.

We adopt $\bm{M}\parallel\hat{\bm{z}}$ throughout. For monoclinic classes ($2$, $2/m$, $2'/m'$, $m'$), the unique monoclinic axis (the twofold rotation axis or, equivalently, the normal mirror plane) is taken along $\hat{\bm{y}}$. The notation used for the symmetry operations in Table~\ref{tab:S3mpg} is as follows.

\begin{itemize}
\item $E$ is the identity operation.
\item $\mathcal{I}$ is the spatial inversion, $(x,y,z) \to (-x,-y,-z)$.
\item $\mathcal{T}$ is the time-reversal operation, which reverses spin and momentum and flips the sign of any $\mathcal{T}$-odd response.
\item $C_{n\alpha}$ denotes the proper rotation by $2\pi/n$ about the axis $\alpha$. Specifically, $C_{2x}$, $C_{2y}$, $C_{2z}$ are twofold rotations about the Cartesian $x$, $y$, $z$ axes; $C_{3z}$ and $C_{3z}^{-1}$ are threefold rotations about $\hat{\bm{z}}$ by $+2\pi/3$ and $-2\pi/3$; $C_{4z}$ and $C_{4z}^{-1}$ are fourfold rotations about $\hat{\bm{z}}$ by $+\pi/2$ and $-\pi/2$; $C_{6z}$ and $C_{6z}^{-1}$ are sixfold rotations about $\hat{\bm{z}}$ by $+\pi/3$ and $-\pi/3$.
\item $m_{\alpha}$ denotes a mirror plane perpendicular to the axis $\alpha$. Specifically, $m_{x}$, $m_{y}$, $m_{z}$ are mirrors perpendicular to $\hat{\bm{x}}$, $\hat{\bm{y}}$, $\hat{\bm{z}}$.
\item $C_{2[\phi]}$ and $m_{[\phi]}$ denote in-plane operations whose axis lies in the $xy$ plane and makes an angle $\phi$ (in degrees) with $\hat{\bm{x}}$. $C_{2[\phi]}$ is a twofold rotation about that in-plane axis, and $m_{[\phi]}$ is a mirror whose normal is the same in-plane axis.
\item $S_{n}$ denotes a rotoinversion (improper rotation): $S_{4z} = m_{z}\,C_{4z}$ and $S_{4z}^{-1} = m_{z}\,C_{4z}^{-1}$ are improper fourfold operations; $S_{6} = \mathcal{I}\,C_{3z}$ and $S_{6}^{-1} = \mathcal{I}\,C_{3z}^{-1}$ are improper sixfold operations associated with the $\bar{3}$ axis; $S_{3} = m_{z}\,C_{3z}$ and $S_{3}^{-1} = m_{z}\,C_{3z}^{-1}$ are improper threefold operations.
\item A prefactor $\mathcal{T}$ in front of any spatial operation $g$ (as in $\mathcal{T} C_{2x}$, $\mathcal{T} m_{y}$, etc.) denotes the antiunitary operation $\mathcal{T}g$, in which the spatial operation $g$ is combined with time reversal. Such operations appear only in the coset $\mathcal{T}\!\cdot\!A$ of type-III MPGs.
\end{itemize}

These generators are precisely those used in our numerical implementation, so the canonical tensor patterns listed in Table~\ref{tab:S3mpg} are directly reproduced by the linear-constraint procedure of Sect.~\ref{S2:implementation}.

\begin{table}[h]
\centering
\scriptsize
\renewcommand{\arraystretch}{2}
\setlength{\tabcolsep}{4pt}
\caption{lists the operation sets or generating sets used to construct $H$ and $\mathcal{T}\!\cdot\!A$.}
\label{tab:S3mpg}
\begin{tabular}{l c c c p{0.26\textwidth} p{0.28\textwidth} c c}
\hline\hline
MPG & MSG type & $|H|$ & $|A|$ &
$H$ & $\mathcal{T}\!\cdot\!A$ &
$\mathcal{T}$-even & $\mathcal{T}$-odd \\
\hline
$\bar{1}$    & I   & 2  & 0  &
$E,\,\mathcal{I}$ &
--- &
$\mathcal{P}_{1}$ & $\mathcal{P}_{1}$ \\

$2$          & I   & 2  & 0  &
$E,\,C_{2y}$ &
--- &
$\mathcal{P}_{3}$ & $\mathcal{P}_{3}$ \\

$2/m$        & I   & 4  & 0  &
$E,\,C_{2y},\,\mathcal{I},$ \newline
$m_{y}$ &
--- &
$\mathcal{P}_{3}$ & $\mathcal{P}_{3}$ \\

$mm2$        & I   & 4  & 0  &
$E,\,C_{2z},\,m_{x},$ \newline
$m_{y}$ &
--- &
$\mathcal{P}_{2}$ & $\mathcal{P}_{2}$ \\

$mmm$        & I   & 8  & 0  &
$E,\,C_{2x},\,C_{2y},$ \newline
$C_{2z},\,\mathcal{I},\,m_{x},$ \newline
$m_{y},\,m_{z}$ &
--- &
$\mathcal{P}_{2}$ & $\mathcal{P}_{2}$ \\

$\bar{4}2m$  & I   & 8  & 0  &
$E,\,C_{2z},\,S_{4z},$ \newline
$S_{4z}^{-1},\,C_{2x},\,C_{2y},$ \newline
$m_{[45]},\,m_{[135]}$ &
--- &
$\mathcal{P}_{7}$ & $\mathcal{P}_{7}$ \\

$4/mmm$      & I   & 16 & 0  &
closure of \newline
$\{C_{4z},\,C_{2x},\,\mathcal{I}\}$ &
--- &
$\mathcal{P}_{7}$ & $\mathcal{P}_{7}$ \\

$3m$         & I   & 6  & 0  &
$E,\,C_{3z},\,C_{3z}^{-1},$ \newline
$m_{[90]},\,m_{[210]},\,m_{[330]}$ &
--- &
$\mathcal{P}_{11}$ & $\mathcal{P}_{11}$ \\

$\bar{3}m$   & I   & 12 & 0  &
$E,\,C_{3z},\,C_{3z}^{-1},$ \newline
$\mathcal{I},\,S_{6},\,S_{6}^{-1},$ \newline
$C_{2[0]},\,C_{2[120]},\,C_{2[240]},$ \newline
$m_{[0]},\,m_{[120]},\,m_{[240]}$ &
--- &
$\mathcal{P}_{10}$ & $\mathcal{P}_{10}$ \\

\hline

$m'$         & III & 1  & 1  &
$E$ &
$\mathcal{T} m_{y}$ &
$\mathcal{P}_{3}$ & $\mathcal{P}_{4}$ \\

$2'/m'$      & III & 2  & 2  &
$E,\,\mathcal{I}$ &
$\mathcal{T} C_{2y},\,\mathcal{T} m_{y}$ &
$\mathcal{P}_{3}$ & $\mathcal{P}_{4}$ \\

$2'2'2$      & III & 2  & 2  &
$E,\,C_{2z}$ &
$\mathcal{T} C_{2x},\,\mathcal{T} C_{2y}$ &
$\mathcal{P}_{2}$ & $\mathcal{P}_{5}$ \\

$m'm2'$      & III & 2  & 2  &
$E,\,m_{y}$ &
$\mathcal{T} C_{2z},\,\mathcal{T} m_{x}$ &
$\mathcal{P}_{2}$ & $\mathcal{P}_{6}$ \\

$m'm'2$      & III & 2  & 2  &
$E,\,C_{2z}$ &
$\mathcal{T} m_{x},\,\mathcal{T} m_{y}$ &
$\mathcal{P}_{2}$ & $\mathcal{P}_{5}$ \\

$m'm'm$      & III & 4  & 4  &
$E,\,C_{2z},\,\mathcal{I},$ \newline
$m_{z}$ &
$\mathcal{T} C_{2x},\,\mathcal{T} C_{2y},\,\mathcal{T} m_{x},$ \newline
$\mathcal{T} m_{y}$ &
$\mathcal{P}_{2}$ & $\mathcal{P}_{5}$ \\

$4'/mm'm$    & III & 8  & 8  &
$E,\,C_{2x},\,C_{2y},$ \newline
$C_{2z},\,\mathcal{I},\,m_{x},$ \newline
$m_{y},\,m_{z}$ &
$\mathcal{T} C_{4z},\,\mathcal{T} C_{4z}^{-1},\,\mathcal{T} S_{4z},$ \newline
$\mathcal{T} S_{4z}^{-1},\,\mathcal{T} C_{2[45]},\,\mathcal{T} C_{2[135]},$ \newline
$\mathcal{T} m_{[45]},\,\mathcal{T} m_{[135]}$ &
$\mathcal{P}_{7}$ & $\mathcal{P}_{8}$ \\

$6'/m'mm'$   & III & 12 & 12 &
$E,\,C_{3z},\,C_{3z}^{-1},$ \newline
$\mathcal{I},\,S_{6},\,S_{6}^{-1},$ \newline
$C_{2[30]},\,C_{2[90]},\,C_{2[150]},$ \newline
$m_{[30]},\,m_{[90]},\,m_{[150]}$ &
$\mathcal{T} C_{6z},\,\mathcal{T} C_{6z}^{-1},\,\mathcal{T} C_{2z},$ \newline
$\mathcal{T} S_{3},\,\mathcal{T} S_{3}^{-1},\,\mathcal{T} m_{z},$ \newline
$\mathcal{T} C_{2[0]},\,\mathcal{T} C_{2[60]},\,\mathcal{T} C_{2[120]},$ \newline
$\mathcal{T} m_{[0]},\,\mathcal{T} m_{[60]},\,\mathcal{T} m_{[120]}$ &
$\mathcal{P}_{7}$ & $\mathcal{P}_{9}$ \\
\hline\hline
\end{tabular}
\end{table}

\subsection{Material-level mapping}

The material-level mapping is summarized in Tables~\ref{tab:S3materials1} and \ref{tab:S3materials2}. Since the tensor forms are already defined at the pattern level, the material table reports only the MPG, the MSG type, the corresponding canonical tensor patterns and the allowed/independent component counts.

\begin{table}[p]
\centering
\scriptsize
\renewcommand{\arraystretch}{1.08}
\setlength{\tabcolsep}{3pt}
\caption{Material-level mapping for entries 1--31 of the Guo altermagnet list.
Metallic compounds are marked by $^{*}$, following Ref.~\cite{spin_split}. When the same chemical formula appears more than once, the parenthesized number denotes the corresponding entry number in the Guo list.}
\label{tab:S3materials1}
\begin{tabular}{c p{0.27\textwidth} c c c c c}
\hline\hline
No. & Compound & MPG & Type &
$\mathcal{T}$-even & $\mathcal{T}$-odd & Count even/odd \\
\hline
1  & BaCrF$_5$ & $2'2'2$ & III & $\mathcal{P}_{2}$ & $\mathcal{P}_{5}$ & 6/6,\;7/7 \\
2  & LaCrO$_3$ (\#2) & $mmm$ & I & $\mathcal{P}_{2}$ & $\mathcal{P}_{2}$ & 6/6,\;6/6 \\
3  & ScCrO$_3$ & $mmm$ & I & $\mathcal{P}_{2}$ & $\mathcal{P}_{2}$ & 6/6,\;6/6 \\
4  & InCrO$_3$ & $mmm$ & I & $\mathcal{P}_{2}$ & $\mathcal{P}_{2}$ & 6/6,\;6/6 \\
5  & TlCrO$_3$ & $mmm$ & I & $\mathcal{P}_{2}$ & $\mathcal{P}_{2}$ & 6/6,\;6/6 \\
6  & TbCrO$_3$ & $m'm'm$ & III & $\mathcal{P}_{2}$ & $\mathcal{P}_{5}$ & 6/6,\;7/7 \\
7  & LaCrO$_3$ (\#7) & $m'm'm$ & III & $\mathcal{P}_{2}$ & $\mathcal{P}_{5}$ & 6/6,\;7/7 \\
8  & YCrO$_3$ & $m'm'm$ & III & $\mathcal{P}_{2}$ & $\mathcal{P}_{5}$ & 6/6,\;7/7 \\
9  & CrNb$_4$S$_8^{*}$ & $6'/m'mm'$ & III & $\mathcal{P}_{7}$ & $\mathcal{P}_{9}$ & 6/3,\;4/1 \\
10 & CrSb$^{*}$ & $6'/m'mm'$ & III & $\mathcal{P}_{7}$ & $\mathcal{P}_{9}$ & 6/3,\;4/1 \\
11 & RbMnF$_4$ & $\bar{1}$ & I & $\mathcal{P}_{1}$ & $\mathcal{P}_{1}$ & 27/27,\;27/27 \\
12 & MnTiO$_3$ & $m'$ & III & $\mathcal{P}_{3}$ & $\mathcal{P}_{4}$ & 13/13,\;14/14 \\
13 & MnCO$_3$ & $2/m$ & I & $\mathcal{P}_{3}$ & $\mathcal{P}_{3}$ & 13/13,\;13/13 \\
14 & Ca$_3$Mn$_2$O$_7$ & $m'm2'$ & III & $\mathcal{P}_{2}$ & $\mathcal{P}_{6}$ & 6/6,\;7/7 \\
15 & SrMn$_2$V$_2$O$_8$ & $m'm2'$ & III & $\mathcal{P}_{2}$ & $\mathcal{P}_{6}$ & 6/6,\;7/7 \\
16 & Mn(N(CN)$_2$)$_2$ & $m'm'm$ & III & $\mathcal{P}_{2}$ & $\mathcal{P}_{5}$ & 6/6,\;7/7 \\
17 & LaMnO$_3$ & $m'm'm$ & III & $\mathcal{P}_{2}$ & $\mathcal{P}_{5}$ & 6/6,\;7/7 \\
18 & Mn$_2$SeO$_3$F$_2$ & $m'm'm$ & III & $\mathcal{P}_{2}$ & $\mathcal{P}_{5}$ & 6/6,\;7/7 \\
19 & Ba$_2$MnSi$_2$O$_7$ & $\bar{4}2m$ & I & $\mathcal{P}_{7}$ & $\mathcal{P}_{7}$ & 6/3,\;6/3 \\
20 & ZrMn$_2$Ge$_4$O$_{12}$ & $4'/mm'm$ & III & $\mathcal{P}_{7}$ & $\mathcal{P}_{8}$ & 6/3,\;6/3 \\
21 & MnF$_2$ & $4'/mm'm$ & III & $\mathcal{P}_{7}$ & $\mathcal{P}_{8}$ & 6/3,\;6/3 \\
22 & KMnF$_3$ & $4/mmm$ & I & $\mathcal{P}_{7}$ & $\mathcal{P}_{7}$ & 6/3,\;6/3 \\
23 & Fe$_2$O$_3$ (\#23) & $\bar{1}$ & I & $\mathcal{P}_{1}$ & $\mathcal{P}_{1}$ & 27/27,\;27/27 \\
24 & LiFeP$_2$O$_7$ & $2$ & I & $\mathcal{P}_{3}$ & $\mathcal{P}_{3}$ & 13/13,\;13/13 \\
25 & Fe$_3$(PO$_4$)$_2$(OH)$_2$ & $2/m$ & I & $\mathcal{P}_{3}$ & $\mathcal{P}_{3}$ & 13/13,\;13/13 \\
26 & FeSO$_4$F & $2'/m'$ & III & $\mathcal{P}_{3}$ & $\mathcal{P}_{4}$ & 13/13,\;14/14 \\
27 & FeOHSO$_4$ & $2'/m'$ & III & $\mathcal{P}_{3}$ & $\mathcal{P}_{4}$ & 13/13,\;14/14 \\
28 & FeF$_3$ & $2'/m'$ & III & $\mathcal{P}_{3}$ & $\mathcal{P}_{4}$ & 13/13,\;14/14 \\
29 & FeBO$_3$ & $2'/m'$ & III & $\mathcal{P}_{3}$ & $\mathcal{P}_{4}$ & 13/13,\;14/14 \\
30 & Fe$_2$O$_3$ (\#30) & $2'/m'$ & III & $\mathcal{P}_{3}$ & $\mathcal{P}_{4}$ & 13/13,\;14/14 \\
31 & ZnFeF$_5$(H$_2$O)$_2$ & $mm2$ & I & $\mathcal{P}_{2}$ & $\mathcal{P}_{2}$ & 6/6,\;6/6 \\
\hline\hline
\end{tabular}
\end{table}

\begin{table}[p]
\centering
\scriptsize
\renewcommand{\arraystretch}{1.08}
\setlength{\tabcolsep}{3pt}
\caption{Material-level mapping for entries 32--62 of the Guo altermagnet list.}
\label{tab:S3materials2}
\begin{tabular}{c p{0.27\textwidth} c c c c c}
\hline\hline
No. & Compound & MPG & Type &
$\mathcal{T}$-even & $\mathcal{T}$-odd & Count even/odd \\
\hline
32 & Fe$_2$PO$_5$ & $mmm$ & I & $\mathcal{P}_{2}$ & $\mathcal{P}_{2}$ & 6/6,\;6/6 \\
33 & SmFeO$_3$ (\#33) & $m'm'm$ & III & $\mathcal{P}_{2}$ & $\mathcal{P}_{5}$ & 6/6,\;7/7 \\
34 & NdFeO$_3$ & $m'm'm$ & III & $\mathcal{P}_{2}$ & $\mathcal{P}_{5}$ & 6/6,\;7/7 \\
35 & TbFeO$_3$ & $m'm'm$ & III & $\mathcal{P}_{2}$ & $\mathcal{P}_{5}$ & 6/6,\;7/7 \\
36 & SmFeO$_3$ (\#36) & $m'm'm$ & III & $\mathcal{P}_{2}$ & $\mathcal{P}_{5}$ & 6/6,\;7/7 \\
37 & Sr$_4$Fe$_4$O$_{11}$ & $m'm'm$ & III & $\mathcal{P}_{2}$ & $\mathcal{P}_{5}$ & 6/6,\;7/7 \\
38 & LiFe$_2$F$_6$ & $4'/mm'm$ & III & $\mathcal{P}_{7}$ & $\mathcal{P}_{8}$ & 6/3,\;6/3 \\
39 & FeCO$_3$ & $\bar{3}m$ & I & $\mathcal{P}_{10}$ & $\mathcal{P}_{10}$ & 10/4,\;10/4 \\
40 & Sr$_2$CoTeO$_6$ & $2/m$ & I & $\mathcal{P}_{3}$ & $\mathcal{P}_{3}$ & 13/13,\;13/13 \\
41 & Li$_2$Co(SO$_4$)$_2$ & $2'/m'$ & III & $\mathcal{P}_{3}$ & $\mathcal{P}_{4}$ & 13/13,\;14/14 \\
42 & Ba$_2$CoGe$_2$O$_7$ & $m'm2'$ & III & $\mathcal{P}_{2}$ & $\mathcal{P}_{6}$ & 6/6,\;7/7 \\
43 & CoF$_2$ & $4'/mm'm$ & III & $\mathcal{P}_{7}$ & $\mathcal{P}_{8}$ & 6/3,\;6/3 \\
44 & CoF$_3$ & $\bar{3}m$ & I & $\mathcal{P}_{10}$ & $\mathcal{P}_{10}$ & 10/4,\;10/4 \\
45 & NiCO$_3$ & $2/m$ & I & $\mathcal{P}_{3}$ & $\mathcal{P}_{3}$ & 13/13,\;13/13 \\
46 & La$_2$NiO$_4$ & $m'm'm$ & III & $\mathcal{P}_{2}$ & $\mathcal{P}_{5}$ & 6/6,\;7/7 \\
47 & NiF$_2$ & $m'm'm$ & III & $\mathcal{P}_{2}$ & $\mathcal{P}_{5}$ & 6/6,\;7/7 \\
48 & NiFePO$_5$ & $mmm$ & I & $\mathcal{P}_{2}$ & $\mathcal{P}_{2}$ & 6/6,\;6/6 \\
49 & PbNiO$_3$ & $3m$ & I & $\mathcal{P}_{11}$ & $\mathcal{P}_{11}$ & 10/4,\;10/4 \\
50 & Y$_2$Cu$_2$O$_5$ & $mm2$ & I & $\mathcal{P}_{2}$ & $\mathcal{P}_{2}$ & 6/6,\;6/6 \\
51 & Cu$_2$V$_2$O$_7$ & $m'm'2$ & III & $\mathcal{P}_{2}$ & $\mathcal{P}_{5}$ & 6/6,\;7/7 \\
52 & CuFePO$_5$ & $mmm$ & I & $\mathcal{P}_{2}$ & $\mathcal{P}_{2}$ & 6/6,\;6/6 \\
53 & Ca$_3$LiRuO$_6$ & $2'/m'$ & III & $\mathcal{P}_{3}$ & $\mathcal{P}_{4}$ & 13/13,\;14/14 \\
54 & Sr$_3$LiRuO$_6$ & $2'/m'$ & III & $\mathcal{P}_{3}$ & $\mathcal{P}_{4}$ & 13/13,\;14/14 \\
55 & RuO$_2^{*}$ & $4'/mm'm$ & III & $\mathcal{P}_{7}$ & $\mathcal{P}_{8}$ & 6/3,\;6/3 \\
56 & Ba$_3$NiRu$_2$O$_9$ & $6'/m'mm'$ & III & $\mathcal{P}_{7}$ & $\mathcal{P}_{9}$ & 6/3,\;4/1 \\
57 & K$_2$ReI$_6$ & $2/m$ & I & $\mathcal{P}_{3}$ & $\mathcal{P}_{3}$ & 13/13,\;13/13 \\
58 & Sr$_2$CoOsO$_6$ & $2/m$ & I & $\mathcal{P}_{3}$ & $\mathcal{P}_{3}$ & 13/13,\;13/13 \\
59 & Ca$_3$LiOsO$_6$ & $2'/m'$ & III & $\mathcal{P}_{3}$ & $\mathcal{P}_{4}$ & 13/13,\;14/14 \\
60 & NaOsO$_3$ & $m'm'm$ & III & $\mathcal{P}_{2}$ & $\mathcal{P}_{5}$ & 6/6,\;7/7 \\
61 & Ba$_3$CoIr$_2$O$_9$ & $2/m$ & I & $\mathcal{P}_{3}$ & $\mathcal{P}_{3}$ & 13/13,\;13/13 \\
62 & CaIrO$_3$ & $m'm'm$ & III & $\mathcal{P}_{2}$ & $\mathcal{P}_{5}$ & 6/6,\;7/7 \\
\hline\hline
\end{tabular}
\end{table}

The survey confirms that the separation between the two SHC channels is not specific to the six representative materials discussed in the main text. For type-I MPGs, the antiunitary coset is absent, and therefore the $\mathcal{T}$-even and $\mathcal{T}$-odd tensor spaces have the same canonical tensor pattern. This is seen, for example, in the entries $mmm$, $2/m$, $4/mmm$, $\bar{3}m$, and $3m$. By contrast, type-III MPGs can distinguish the two channels because operations in the antiunitary coset act with opposite signs in $\mathcal{T}$-even and $\mathcal{T}$-odd responses. This produces distinct tensor patterns in $2'/m'$, $m'$, $m'm'm$, $m'm2'$, $4'/mm'm$, and $6'/m'mm'$.

Several trends are apparent at the MPG level. Low-symmetry centrosymmetric classes such as $\bar{1}$ leave all 27 SHC components allowed. Monoclinic classes such as $2/m$ and $2'/m'$ allow large tensor spaces with 13 or 14 independent components depending on the time-reversal parity. Higher rotational symmetries strongly reduce the tensor rank. The tetragonal and hexagonal patterns contain only six allowed $\mathcal{T}$-even components with three independent parameters, while the $\mathcal{T}$-odd sector of $6'/m'mm'$ is reduced to four allowed components with only one independent parameter. Thus, the Guo-list extension provides a symmetry prescreening map for identifying which MPG classes can host a desired SHC tensor sector before performing dense first-principles transport calculations.

\clearpage
\bibliographystyle{apsrev4-2}
\bibliography{biblio}

\end{document}